# Key Generation Using External Source Excitation: Capacity, Reliability, and Secrecy Exponent

Tzu-Han Chou, *Student Member, IEEE,* Stark C. Draper, *Member, IEEE,* and Akbar M. Sayeed, *Senior Member, IEEE.*

***Abstract*—We study the fundamental limits to secret key generation from an *excited distributed source* (EDS). In an EDS a pair of terminals observe dependent sources of randomness excited by a pre-arranged signal. We first determine the secret key capacity for such systems with one-way public messaging. We then characterize a tradeoff between the secret key rate and exponential bounds on the probability of key agreement failure and on the secrecy of the key generated. We find that there is a fundamental tradeoff between reliability and secrecy.**

**We then explore this framework within the context of reciprocal wireless channels. In this setting, the users transmit pre-arranged excitation signals to each other. When the fading is Rayleigh, the observations of the users are jointly Gaussian sources. We show that an on-off excitation signal with an SNR-dependent duty cycle achieves the secret key capacity of this system. Furthermore, we characterize a fundamental metric – minimum energy per key bit for reliable key generation – and show that in contrast to conventional AWGN channels, there is a non-zero threshold SNR that achieves the minimum energy per key bit. The capacity achieving on-off excitation signal achieves the minimum energy per key bit at any SNR below the threshold. Finally, we build off our error exponent results to investigate the energy required to generate a key using a finite block length. Again we find that on-off excitation signals yield an improvement when compared to constant excitation signals. In addition to Rayleigh fading, we analyze the performance of a system based on binary channel phase quantization.**

## I. Introduction

In [3] Shannon laid the theoretical foundations of cryptography. He defined the notion of *perfect secrecy* achieved by a system wherein the posterior probability of the plain-text message conditioned on an eavesdropper's knowledge is equal to the priori distribution. In other words, the eavesdropper can deduce no information about the plain-text based on his (or her) observations. Shannon also showed that, in order to achieve perfect secrecy, a secret key at least as large as the message must be shared by encrypter and decrypter. Due to the difficulty of distributing secret keys securely, much of cryptography has followed a distinct philosophy where the security is based on the computational hardness of certain mathematical problems. This is termed *computational security*.

Starting in the 1970s, information theorists followed up on Shannon's work by exploring security systems that exploited auxiliary sources of randomness to facilitate key generation. Since our work is motivated by considering auxiliary sources based on channel randomness, we refer to this body of work as *physical-layer* secrecy. The earliest results along these lines came from Wyner who in [4] introduces the wire-tap channel. Wyner's work is generalized by Csiszár and Körner in [5]. These results show that, given statistical knowledge of communication channels that link a transmitter to an intended receiver and to an eavesdropper, one may design a code such that messages can be successfully and confidentially sent without the transmitter and intended receiver having access to a shared key.

Later, first by Maurer in [6] and then by Ahlswede and Csiszár in [7], a different use of auxiliary randomness was posed. In these papers a secret key is generated from dependent auxiliary sources of randomness available to two users. The resulting key can then be used, as in Shannon's original architecture, as a one-time pad. In this setting two legitimate users, Alice and Bob, observe dependent discrete memoryless sources (DMS) $X_a^n$ and $X_b^n$. An eavesdropper Eve observes a third statistically related source $X_e^n$. Based on $n$ observations of the sources and a public conversation between Alice and Bob, Alice and Bob want to generated a shared secret key. The public messages $\Phi$ that make up Alice and Bob's conversation must not leak information of the resulting key to Eve. The largest achievable secret key rate (bits per source symbol) is termed the *secret key capacity*. One possible implementation is for Alice to send a single message to Bob (and to have Bob send no messages). The public message indexes a subset of all possible source sequences Alice might have observed, and in which her observed sequence lies. Using his observation and public message, Bob can determine with high probability which sequence Alice observed, while Eve is left with some ambiguity. The shared key can then be generated from the determined sequence. Cast in this setting, the key generation problem can be viewed as a variant of distributed source coding with side information [8]. Reliable recovery of $X_a^n$ by Bob requires the rate of the public message to be greater than $H(X_a|X_b)$.

Manuscript received November 1, 2010; revised April 13, 2011. This work has been supported in part by the National Science Foundation under CAREER Grant No. CF-0844539 and Grant No. CNS-0627589. This work was presented in part at the UCSD Workshop on Inform. Theory Applications, San Diego, CA January 2009 [1] and in part at the at the IEEE Int. Symp. Inform. Theory, Seoul, Korea, June 2009 [2].

The authors are with the Dept. of Electrical and Computer Engineering, University of Wisconsin, Madison, WI 53706. E-mail: tchou2@wisc.edu, sdraper@ece.wisc.edu, akbar@engr.wisc.edu.

In wireless communications, the channel itself can be used as the source of auxiliary randomness. In this case the randomness is due to multipath fading [9]. When transmissions are bi-directional and in the same frequency band (e.g., time-division duplexed systems), the reciprocity property [10] of electromagnetic wave propagation ensures that the channel in each direction is identical. If two channel soundings, one for each direction, are done within the coherence time, the two users' observations are highly correlated. In this setting, $X_a, X_b, X_e$ can be generated by having each user transmit pre-arranged sounding signals. This motivates the general study of key generation from *excited* sources of randomness where the designer chooses the source of excitation. Returning to the wireless setting, the keys $K_a$ and $K_b$, are functions of the estimates of the common random channel observed by each user. For wideband and multi-antenna channels, the resulting keys can be quite large due to the large number of independent channel degrees of freedom. Secrecy from eavesdroppers is ensured by the physics of propagation. Consider, for instance, a rich multipath environment. If the eavesdropper is physically displaced from the legitimate users, by even a few wavelengths, its channel output $X_e$ will nearly be statistically independent of $X_a$ and $X_b$. For these reasons, reciprocal wireless channels provide an attractive source of randomness for key generation.

Motivated by the above discussion of wireless channels, in this paper we study secret key generation from *excited distributed source* (EDS) in which the shared randomness between two parties is excited by a design signal. We term such a source as an *excited distributed memoryless source* (EDMS) if the source is memoryless. We show that the secret key capacity of a general EDMS is an extension of secret key capacity of DMS in the following sense: for a fixed excitation signal $S = s$ the excited source $(X_a, X_b, X_e)$ is a DMS and has secret key capacity $C_{\rm K}(s)$. Thus $s_i$ can be viewed as the *state* of the system at time $i$. For a *known* excitation sequence $s^n$ with type $P_{s^n}$, we achieve secret key rate $\sum_s P_{s^n}(s)C_{\rm K}(s)$ by time sharing among the DMSs. Since $s^n$ is a design choice, and the set of types is dense in the set of all distributions, the secret key capacity of an EDMS is obtained by maximizing the secret key rate over all possible excitation distributions $p_S$, perhaps under some input cost constraint (such as power). This result is specialized to give the secret key capacity of an EDMS with a degraded eavesdropper. We also characterize a reliability-secrecy tradeoff by quantifying two exponents. The first exponent $E_{\rm R}$ bounds the probability that two users cannot reconcile the same key, i.e., $-\frac{1}{n}\log \Pr(K_a \neq K_b) \geq E_{\rm R} + o(1)$. The second exponent $E_{\rm S}$ bounds the information leakage to the eavesdropper, characterized by $-\frac{1}{n}\log I(K_a; X_e^n, S^n, \Phi) \geq E_{\rm S} + o(1)$. The overall result yields a strongly achievable secret key rate-exponent triple $(R_{\rm SK}, E_{\rm R}, E_{\rm S})$. For a given $R_{\rm SK}$ less than capacity, there is a tradeoff between achievable $E_{\rm R}$ and $E_{\rm S}$. An example of such a tradeoff is shown in Figure 1. For a set rate, the tradeoff between the two exponents is achieved by varying the rate of the public message.

We then apply our results on EDMSs to a reciprocal Rayleigh fading channel. We discuss the limiting cases of un-quantized Gaussian sources as well as binary sources that result from quantizing uniform channel phases. We show that the choice of excitation signal has a significant effect on the possible rates of key generation. In the case where the excitation signal is subject to a power constraint, a signal that uniformly distributes power across all channels degrees of freedom results in a secret key rate that is convex in low-SNR (signal-to-noise ratio) and concave in high-SNR. Due to the convexity in low-SNR, uniform excitation signals are not always optimal. We show that an on-off excitation signal which uses only a fraction of the available degrees of freedom (but at a higher SNR) is the optimal input distribution at low SNR and characterize the secret key capacity at all SNRs. Our analysis of Gaussian sources is based on an equivalent Gaussian noise model. This facilitates comparison with the conventional additive white Gaussian noise (AWGN) channel.

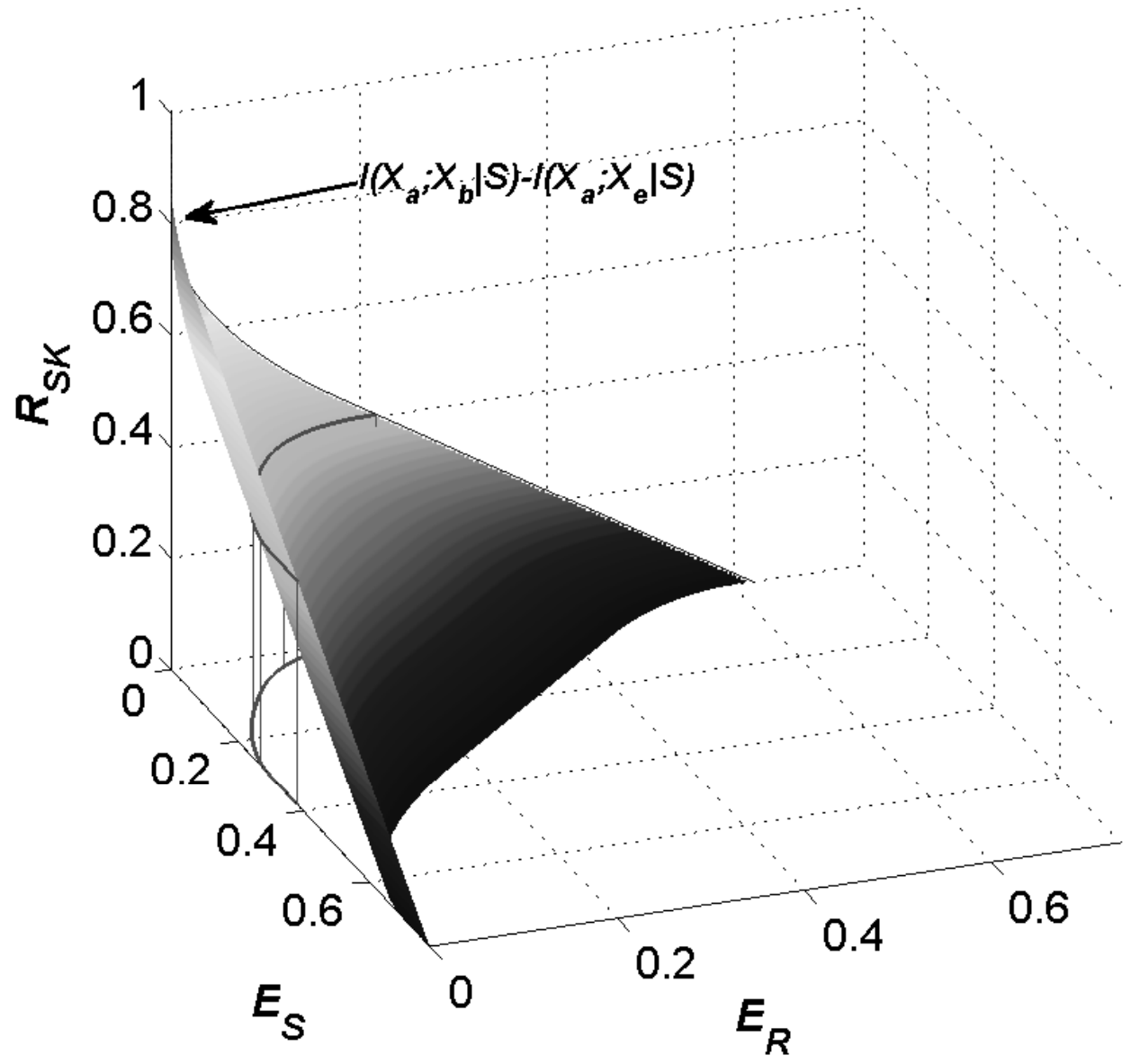

Fig. 1. Achievable $(R_{\rm SK}, E_{\rm R}, E_{\rm S})$

The behavior of the secret key capacity in the low-SNR regime also reveals a fundamental quantity, the *minimum energy per key bit*. This quantity is reciprocally related to the capacity per unit cost [11] of the system. We show that, unlike the minimum energy per bit for reliable communication in an AWGN channel, there is a non-zero threshold SNR at which the minimum energy per key bit for our system is achieved. At lower SNRs with the optimal on-off signaling, the key capacity-cost function has the same slope as at the threshold SNR. This means it yields the same minimum energy per key bit. Finally, we characterize the minimum energy required for reliable key generation with a finite block length. We give a lower bound on the energy required by building off our results on the reliability exponent.

## A. Related work

As mentioned above, key generation from general correlated sources was originally studied in [6, 7]. Those papers introduce the basic model and characterize the secret key capacity. Regarding correlated Gaussian sources (which would be the case if the wireless transmissions experience Rayleigh fading) [12] and [13] investigate key generation from jointly Gaussian variables where [12] uses LDPC coding and [13] uses nested lattice codes and vector quantization for reconciliation. Generation of secret keys based on the inherent randomness of wireless channel is studied in [14–20]. Early work regarding key extraction from reciprocal channel randomness did not consider the use of a public message. For example, in [14] the phase difference between two orthogonal sinusoids in a received signal is quantized and coding is applied to improve the probability of key acquisition. An uncoded protocol is considered in [15] which analyzes the energy resource required for key acquisition. Regarding the approaches that use public discussions and fall within the information-theoretic approaches, [16] and [17] exploit the random amplitude of time delay channels to generate a key where [16] studies the ultra-wideband (UWB) channel and [17] focuses on narrow band channels. In [18] the use of randomness from level crossing process is considered. Adaptive quantization of noisy channel outputs for key generation is investigated in [19] and [20]. There are also related works on secret key generation in which the key is not generated from channel randomness. In [21] a secret key agreement protocol is studied but the key itself is a pseudo-random sequence generated by the user and does not result from the inherent randomness of wireless channels. In [22] key generation is considered in a situation where in addition to the dependent sources, there is a wiretap channel available for use.

### B. Paper outline

This paper is organized as follows. In Section II, we define a general EDMS model and a Gaussian EDMS which results from exciting the randomness of Rayleigh fading channel. We introduce a one-way secret key generation system and state the performance measure considered in this paper. Our main results are summarized in Section III. In Section IV, we discuss the reliability-secrecy tradeoff of the system. For a Gaussian source, we show that an on-off excitation signal is a capacity achieving signal in low-SNR and discuss reliability exponent and the energy consumption of a key with a finite block length using an on-off signal. A simple key generation scheme using binary phase quantization is also presented in Section IV. The proofs of the main results are given in Section V. Section VI concludes the paper.

### C. Notation

A random variable is represented by a capital letter such as $X$ and a fixed-value scalar is represented by a lower case letter such as $x$. A sequence $(X_i, X_{i+1}, \cdots, X_j)$ is denoted by $X_i^j$ and a similar notation $x_i^j$ is for a fixed-value sequence. When a sequence starts from $i = 1$, we use $X^j$ (or $x^j$) as a shorthand. Sets are denoted by using a calligraphic font, e.g., $\mathcal{S}$. The complement of a set $\mathcal{S}$ is denoted by $\mathcal{S}^c$.

We will use the notation $p_X(x)$ for the probability distribution function of $X$ for both discrete and continuous cases. This will not cause any confusion: $p_X(x)$ represents a probability mass function when $X$ is a discrete random variable and it represents a probability density function when $X$ is a continuous random variable. We use capital $P_{x^n}(x)$ to denote the type (empirical distribution) of a sequence $x^n$. We also use a shorthand for state dependent distribution, e.g., $Q_s(x) = p_{X|S}(x|s)$ is the conditional distribution of $X$ when state $S = s$. Throughout the paper, $\log(\cdot)$ refers to the natural logarithm and $\log_2(\cdot)$ refers to the logarithm in base two.

## II. Secret Key From excited source: models and definitions

In this section we provide basic definitions and introduce the models with which we work in this paper. We start in Section II-A by defining an excited source of common randomness where the users (Alice and Bob) can influence the source of randomness from which they want to distill a common key. We also define the class of key distillation systems with which we work. In Section II-B we give the performance measures we aim to optimize. In Section II-C we define a particular class of excited sources of common randomness based on reciprocal multipath fading channels. Based on these definitions, we state the main results in Section III.

### A. Source and System Model

The study of secret key generation from common randomness was initiated in [6, 7] where the users have access to a source of randomness specified, in effect, by nature. However, there are many situations where the users themselves can excite some medium to generate the source of randomness. We formalize our model of such situations by defining an *excited* source of common randomness.

*Definition 1:* Let $\mathcal{X}_a$,$\mathcal{X}_b$,$\mathcal{X}_e$ and $\mathcal{S}$ are finite sets. A length-$n$ *excited distributed source* (EDS) of common randomness is specified by an arbitrary conditional distribution $p_{X_a^n,X_b^n,X_e^n|S^n}(x_a^n, x_b^n, x_e^n|s^n)$ on $\mathcal{X}_a^n \times \mathcal{X}_b^n \times \mathcal{X}_e^n \times \mathcal{S}^n$ where the designer can choose $s^n \in \mathcal{S}^n$. We further call this an *excited distributed memoryless source* (EDMS) if

$$p_{X_a^n,X_b^n,X_e^n|S^n}(x_a^n, x_b^n, x_e^n|s^n) = \prod_{i=1}^{n} p_{X_a,X_b,X_e|S}(x_{a,i}, x_{b,i}, x_{e,i}|s_i). \quad (1)$$

Let $\mathcal{P}_S$ be a set of probability mass distributions. An EDMS is said to have input constraint $\mathcal{P}_S$ (or simply $\mathcal{P}_S$-EDMS) if the type $P_{s^n}$ of $s^n$ is constrained to be in $\mathcal{P}_S$ (for every $n$).

We note that since $s^n$ is a system parameter (in effect the "excitation" or "sounding" signal) and is known by all parties – Alice, Bob, and Eve. It will be useful to consider each element of $\mathcal{S}$ as a system state, known to all, upon which the source of randomness is conditioned. Hence we often refer to $s \in \mathcal{S}$ as the system "state".

A special case of source in which Eve (Bob) has degraded observation of Bob (Eve) is defined as follows.

*Definition 2 (Degradedness):* Given an EDMS we say that source $X_e$ is a *degraded* version of $X_b$ when $S = s$ if

$$p_{X_a,X_b,X_e|S}(x_a, x_b, x_e|s) = p_{X_a,X_b|S}(x_a, x_b|s) p_{X_e|X_b,S}(x_e|x_b, s) \ .$$

In other words, if, conditioned on $S = s$, $X_a - X_b - X_e$ forms a Markov chain. An EDMS is termed an *EDMS with degraded states* if for every $S = s$, either $X_e$ is a degraded version of $X_b$ or $X_b$ is a degraded version of $X_e$. Finally, an EDMS has a *degraded eavesdropper* if $X_e$ is a degraded version of $X_b$ for all states.

We consider key generation systems in which Alice, Bob and Eve respectively observe $X_a^n$, $X_b^n$ and $X_e^n$. The excitation signal $S^n$ is known by all. Alice and Bob are allowed to communicated via an error-free public channel that is monitored, but not impeded, by Eve. We concentrate on secret key generation based on one-way (Alice to Bob) public discussions (cf., e.g. [7]) defined next.

*Definition 3:* A *one-way secret key generation system* with excitation signal $s^n$ is defined by a triplet of functions

$$\begin{aligned} f_a(\,\cdot\,; s^n) &: \mathcal{X}_a^n \to \mathcal{K}, \\ g(\,\cdot\,; s^n) &: \mathcal{X}_a^n \to \mathcal{M}, \\ f_b(\,\cdot\,; s^n) &: \mathcal{X}_b^n \times \mathcal{M} \to \mathcal{K}. \end{aligned}$$

Respectively, these three functions define Alice's key, $K_a$, the public message sent by Alice to Bob, $\Phi$, and Bob's key, $K_b$:

$$\begin{aligned} K_a &= f_a(X_a^n; s^n), \\ \Phi &= g(X_a^n; s^n), \\ K_b &= f_b(X_b^n, \Phi; s^n). \end{aligned}$$

### B. Performance Measures

We next consider the performance measures of interest. These include capacity-type measures on secret key generation, error-exponent measures on the reliability of key generation, and guarantees on the secrecy of the generated key.

*Definition 4 (Weak Achievability):* A secret key rate $R_{\mathrm{SK}}$ is $\mathcal{P}_S$*-weakly-achievable* if for any $\epsilon > 0$ there exists an $n_0$ such that for all $n > n_0$, there exists a one-way secret key generation system with excitation sequence $s^n$, $P_{s^n} \in \mathcal{P}_S$, that satisfies

$$\frac{1}{n} H(K_a | S^n = s^n) > \frac{1}{n} \log |\mathcal{K}| - \epsilon, \tag{2}$$

$$\frac{1}{n} H(K_a | S^n = s^n) > R_{\mathrm{SK}} - \epsilon, \tag{3}$$

$$\frac{1}{n} I(K_a; X_e^n, \Phi | S^n = s^n) < \epsilon, \tag{4}$$

$$\Pr(K_a \neq K_b) < \epsilon. \tag{5}$$

We remark that in Definition 4 condition (2) means that the key is almost uniformly distributed across the set $\mathcal{K}$. Conditions (3) and (4) imply $\frac{1}{n} H(K_a | X_e^n, \Phi, S^n = s^n) \geq R_{\mathrm{SK}} - 2\epsilon$, i.e., Eve's observation reveals almost no information about the key.

*Definition 5:* The *secret key capacity* of a one-way secret key generation system is

$$C_{\mathrm{K}} = \sup \left\{ R_{\mathrm{SK}} : R_{\mathrm{SK}} \text{ is } \mathcal{P}_S\text{-weakly-achievable} \right\} .$$

In some practical applications, arbitrary small error probability and key information leakage *rate* as increasing block length are insufficient [23]. In this paper, we further study exponential reliability of key reconciliation and the exponential preservation of secrecy. Generally there will be a tradeoff between the two.

*Definition 6 (Achievable secret key rate-exponent triple):* A secret key rate-exponent triple $(R_{\mathrm{SK}}, E_{\mathrm{R}}, E_{\mathrm{S}}) \in \mathbb{R}_+^3$ is *achievable* if there is a one-way secret key generation system such that $R_{\mathrm{SK}}$ is achievable per Definition 4 while (4) and (5) are replaced by stronger conditions

$$\liminf_{n \to \infty} -\frac{1}{n} \log I(K_a; X_e^n, \Phi | S^n = s^n) \geq E_{\mathrm{S}}, \tag{6}$$

$$\liminf_{n \to \infty} -\frac{1}{n} \log \Pr(K_a \neq K_b) \geq E_{\mathrm{R}}. \tag{7}$$

*Definition 7 (Strong achievability):* A $R_{\mathrm{SK}}$ is strongly achievable if $(R_{\mathrm{SK}}, E_{\mathrm{R}}, E_{\mathrm{S}})$ is achievable for some $E_{\mathrm{R}} > 0$ and $E_{\mathrm{S}} > 0$.

*Definition 8 (Capacity-Reliability-Secrecy Region):* The (secret key) capacity-reliability-secrecy region $\mathcal{R} \subset \mathbb{R}_+^3$ is the closure of the set of achievable secret key rate-exponent triples.

### C. Multipath Channel Models: A Gaussian EDMS

An example of EDMS that we study in depth is based on reciprocal wireless fading channels. Say that Alice and Bob are linked by such a channel. Then, to generate a secret key, Alice and Bob alternately use the channel $n$ times to transmit a known excitation sequence $s^n$ to each other. We assume that the channel coefficients remain static over the two successive soundings by Alice and Bob and are independent to each other times. Under these assumptions, the respective outputs observed by Alice and Bob are

$$X_{a,i} = H_i s_i + N_{a,i}, \tag{8a}$$

$$X_{b,i} = H_i s_i + N_{b,i}, \tag{8b}$$

for $1 \leq i \leq n$, where $\{H_i\}$ are the i.i.d. reciprocal channel coefficients. We study the case of rich multipath Rayleigh fading where $H_i$ is a zero-mean complex Gaussian random variable with variance one, i.e., $\mathcal{CN}\left(0, 1\right)$. The noise at Alice ($N_{a,i}$) and Bob ($N_{b,i}$) are respectively independent $\mathcal{CN}\left(0, \sigma_a^2\right)$ and $\mathcal{CN}\left(0, \sigma_b^2\right)$ random variables. We will constrain the excitation sequence $s^n$ to satisfy an average power constraint $\frac{1}{n}\sum_{i=1}^n |s_i|^2 \leq \mathsf{P}$. The received signal-to-noise ratio (SNR) at Alice (resp. Bob) is defined as $\gamma_a = \mathsf{P}/\sigma_a^2$ (resp. $\gamma_b = \mathsf{P}/\sigma_b^2$).

Eve gets two looks at each channel coefficient, one induced by Alice's excitation signal, the other by Bob's. Respectively, these outputs are $X_{ea,i} = H_{ea,i} s_i + N_{ea,i}$ and $X_{eb,i} = H_{eb,i} s_i + N_{eb,i}$. In a rich multipath environment, $H_{ea,i}$ and $H_{eb,i}$ will be independent of the main channel coefficient $H_i$ due to strong spatial decorrelation between channel coefficients. Thus, Eve's channel observation $X_e$ is not useful for estimating the secret key. Thus, she can only use her knowledge of the excitation signal and the public message to estimate the secret key. This is a special case of EDMS with a degraded eavesdropper. When $S^n = s^n$ with $s_i = s$ a constant for all $i$, this situation specializes to *Model S* in [7].

The correlation between the Gaussian sources $X_a$ and $X_b$ given in (8) depends on the transmission power of the excitation signal. We are interested in the most power efficient regime for key generation. In this case, in addition to secret key capacity, we also characterize another fundamental quantity: *minimum energy per key bit*. We note that in this paper we assume the public channel is error-free and restrict attention to the energy consumption in the excitation signal. The following definition is the reciprocal of the capacity per unit cost defined in [11].

*Definition 9:* The *minimum energy per key bit* is defined to be

$$\left(\frac{\mathcal{E}_{\rm b}}{\sigma^2}\right)_{\min} = \inf_{\gamma>0} \frac{\gamma}{C_{\rm K}(\gamma)} \log 2, \tag{9}$$

where, for simplicity, we assume $\sigma_a^2 = \sigma_b^2 = \sigma^2$, $\gamma = \mathsf{P}/\sigma^2$ is the system SNR, and $C_{\rm K}(\gamma)$ is the secret key capacity as a function of $\gamma$.

## III. Main Results

Given the setup of Section II we are now prepared to state the main results of the paper. Theorem 1 gives the secret key capacity for a general EDMS. Strongly achievable $(R_{\rm SK}, E_{\rm R}, E_{\rm S})$ for a degraded eavesdropper is presented in Theorem 2 and Theorem 3. Theorem 4 and 5 present the results for Gaussian EDMS.

### A. General EDMS

Our first result gives the secret key capacity of a general EDMS with input constraint $\mathcal{P}_S$.

*Theorem 1:* The secret key capacity of an EDMS $(X_a, X_b, X_e, S)$ is

$$C_{\rm K} = \max_{\substack{U,T\\ p_S\in\mathcal{P}_S}} \left[I(T;X_b|U,S) - I(T;X_e|U,S)\right] \ . \tag{10}$$

The maximization is over distribution $p_S \in \mathcal{P}_S$ and auxiliary variables $U, T$ with distribution

$$p_{U,T|S}(u,t|s)p_{X_a|T,S}(x_a|t,s)p_{X_b,X_e|X_a,S}(x_b,x_e|x_a,s) \ . \tag{11}$$

Further, the secret key capacity is bounded from above as

$$C_{\rm K} \leq \max_{p_S\in\mathcal{P}_S} I(X_a;X_b|X_e,S) \ . \tag{12}$$

Theorem 1 can be understood by using the interpretation of the known $s^n$ as a sequence of system states. The secret key capacity (10) is expanded as

$$C_{\rm K} = \max_{p_S\in\mathcal{P}_S} \sum_s p_S(s) \max_{U^s,T^s} \Big[I(T^s;X_b|U^s,S=s) - I(T^s;X_e|U^s,S=s)\Big] \ ,$$

where $U^s, T^s$ are state-dependent auxiliary random variables. For any particular value $S = s$, the source $(X_a, X_b, X_e)$ is a discrete memoryless source (DMS) distributed according to $p_{X_a,X_b,X_e|S}(x_a,x_b,x_e|s)$. The capacity-achieving scheme of [7, Theorem 1] for DMSs can be applied to this system point-wise for each state. The secret key capacity is the weighted sum of state-wise secret key capacity, weighted according to the maximizing $p_S \in \mathcal{P}_S$. Capacity is achieved by selecting a sequence of excitation signals $s^n$ so that $P_{s^n}$ approaches $p_S \in \mathcal{P}_S$ as $n \to \infty$.

Since the designer has control of the excitation signal, the best choice is to choose $p_S(s) \in \mathcal{P}_S$ to maximize the achievable rate. In the special case in which there is no constraint imposed on $p_S$, we have

$$C_{\rm K} = \max_{s\in\mathcal{S}} \max_{U,T} I(T;X_b|U,S=s) - I(T;X_e|U,S=s) \ .$$

Namely, we choose a state that has the largest achievable secret key rate.

### B. EDMS with Degraded Eavesdropper

In the degraded setting, a stronger result follows from Theorem 1.

*Corollary 1:* For an EDMS with degraded states the secret key capacity is

$$C_{\rm K} = \max_{p_S\in\mathcal{P}_S} \sum_{s\in\mathcal{S}} p_S(s)|I(X_a;X_b|S=s) - I(X_a;X_e|S=s)|^+,$$

where $|x|^+ = \max\{x, 0\}$. Further, if the EDMS has a degraded eavesdropper, the secret key capacity is

$$C_{\rm K} = \max_{p_S\in\mathcal{P}_S} I(X_a;X_b|S) - I(X_a;X_e|S). \tag{13}$$

To emphasize the state dependence, in the following, we use notation $Q_s(x_b) = p_{X_b|S}(x_b|s)$ and $\tilde{Q}_s(x_e) = p_{X_e|S}(x_e|s)$ as the conditional distribution of $X_b$ and $X_e$ respectively when input $S = s$ and denote the corresponding channel law as $W_s(x_a|x_b) = p_{X_a|X_b,S}(x_a|x_b,s)$ and $V_s(x_a|x_e) = p_{X_a|X_e,S}(x_a|x_e,s)$.

In order to get a small error probability Alice must transmit enough information so that Bob can recover Alice's key. However, while a larger public message rate increases the reliability of key recovery it also reveals more about $X_a^n$ to Eve. In Theorem 2 we study this tradeoff by characterizing an inner bound of capacity-reliability-secrecy region $\mathcal{R}$. The encoding functions $f_a(\,\cdot\,;s^n)$ and $g(\,\cdot\,;s^n)$ used to prove the theorem are random binning codes defined as follows.

*Definition 10:* An $(n, R)$ *random binning code* for alphabet $\mathcal{X}$ is a random map $f : \mathcal{X}^n \to \mathcal{B} = \{1, 2, \cdots \lfloor e^{nR} \rfloor\}$ in which each $x^n \in \mathcal{X}^n$ is independently and uniformly assigned to an element of $\mathcal{B}$.

Thus, we get the definition of a random binning secret key generation system.

*Definition 11:* An $(n, R_{\rm SK}, R_{\rm M})$ *random binning secret key code* with excitation signal $s^n$ consists of two independent random binning codes:

$$f_a(\,\cdot\,;s^n) : \mathcal{X}_a^n \to \mathcal{K} = \{1, 2, \cdots, \lfloor e^{nR_{\rm SK}} \rfloor\}, \tag{14}$$

$$g(\,\cdot\,;s^n) : \mathcal{X}_a^n \to \mathcal{M} = \{1, 2, \cdots, \lfloor e^{nR_{\rm M}} \rfloor\}. \tag{15}$$

where $f_a(\,\cdot\,;s^n)$ is an $(n, R_{\rm SK})$ random-binning code and $g(\,\cdot\,;s^n)$ is an $(n, R_{\rm M})$ random-binning code.

Before presenting our exponent results, we first define a few quantities. The reader may recognize (16)–(18) as the state-dependent versions of Gallager's source-coding with decoder side information exponents [24]. Quantities (19)–(21) will play the same role in the secrecy exponent. Let

$$E_{\mathrm{R}}(R_{\mathrm{M}}) = \max_{0\le\rho\le 1} \rho R_{\mathrm{M}} - E_0(\rho, p_S) \ , \tag{16}$$

$$E_0(\rho, p_S) = \sum_{s\in S} p_S(s)\tilde{E}_0(\rho, s) \ , \tag{17}$$

$$\tilde{E}_0(\rho, s) = \log\left(\sum_{x_b} Q_s(x_b)\left(\sum_{x_a} W_s(x_a|x_b)^{\frac{1}{1+\rho}}\right)^{1+\rho}\right) . \tag{18}$$

and

$$E_{\mathrm{S}}(R_{\mathrm{M}}, R_{\mathrm{SK}}) = \max_{0\le\alpha\le 1} F_0(\alpha, p_S) - \alpha(R_{\mathrm{M}} + R_{\mathrm{SK}}) \tag{19}$$

$$F_0(\alpha, p_S) = \sum_{s\in S} p_S(s)\tilde{F}_0(\alpha, s) \tag{20}$$

$$\tilde{F}_0(\alpha, s) = -\log\left(\sum_{x_e} \tilde{Q}_s(x_e)\sum_{x_a} V_s(x_a|x_e)^{1+\alpha}\right) \tag{21}$$

We say that a secret key rate-exponent region, $\tilde{\mathcal{R}}(p_S, R_{\mathrm{M}})$, parameterized by excitation distribution $p_S$ and auxiliary rate $R_{\mathrm{M}}$, is achievable if for every point $(R_{\mathrm{SK}}, E_{\mathrm{R}}, E_{\mathrm{S}}) \in \tilde{\mathcal{R}}(p_S, R_M)$ there exists a sequence of one-way secret key generation systems that satisfy the following

$$\lim_{n\to\infty} P_{s^n}(s) = p_S(s) \text{ for all } s \in \mathcal{S}, \tag{22a}$$

$$\lim_{n\to\infty} -\frac{1}{n}\log \Pr(K_a \neq K_b) \ge E_{\mathrm{R}}(R_{\mathrm{M}}), \tag{22b}$$

$$\lim_{n\to\infty} -\frac{1}{n}\log I(K_a; X_e^n, \Phi|S^n = s^n) \ge E_{\mathrm{S}}(R_{\mathrm{M}}, R_{\mathrm{SK}}) \tag{22c}$$

The following theorem gives an inner bound of capacity-reliability-secrecy region.

*Theorem 2:* The union of $\tilde{\mathcal{R}}(p_S, R_{\mathrm{M}})$ over distribution $p_S \in \mathcal{P}_S$ and auxiliary public message rate $R_{\mathrm{M}}$ is an inner bound of the capacity-reliability-secrecy region $\mathcal{R}$, i.e.,

$$\bigcup_{R_{\mathrm{M}}, p_S\in\mathcal{P}_S} \tilde{\mathcal{R}}(p_S, R_{\mathrm{M}}) \subseteq \mathcal{R} \ . \tag{23}$$

When there is no secrecy constraint, the key distillation problem becomes a classic Slepian-Wolf problem [8] for which the reliability exponent $E_{\mathrm{R}}$ (16) is due to Gallager [24]. A slight difference is that, due to the known excitation signal, in (18) the reliability exponent is expressed in terms of the state dependent channel law.

A secrecy condition akin to (6) has been studied in [25, 26] in the setting where Alice and Bob have the same observation, i.e., the special case of Theorem 2 when $X_a = X_b$. In this setting a public discussion is not required. Since reconciliation of $X_a$ and $X_b$ is not required, a tradeoff between reliability and secrecy (e.g., illustrated in Fig. 1) cannot be observed. In this paper we build off that analysis (termed "privacy amplification") to understand the effects of the public discussion and of the excitation signal.

The secrecy exponent of (19)-(21) has a form similar to that of a state-dependent Gallager's exponent, expressed in terms of the sum rate $R_{\mathrm{M}} + R_{\mathrm{SK}}$. The resulting $\tilde{E}_0(\rho, s)$ and $\tilde{F}_0(\alpha, s)$ have properties similar to Gallager's exponent, as is summarized in the following theorem.

*Theorem 3 (Properties of $\tilde{E}_0$ and $\tilde{F}_0$):*

1) $\tilde{E}_0(\rho, s)$ is a non-decreasing non-negative function of $\rho$ for $\rho \ge 0$ and $\tilde{E}_0(0, s) = 0$. Furthermore, $\tilde{E}_0(\rho, s)$ is a convex function of $\rho$ and
$$\left.\frac{\partial\tilde{E}_0(\rho, s)}{\partial\rho}\right|_{\rho=0} = H(X_a|X_b, S = s)$$
2) $\tilde{F}_0(\alpha, s)$ is a non-decreasing non-negative function of $\alpha$ for $\alpha \ge 0$ and $\tilde{F}_0(0, s) = 0$. Furthermore, $\tilde{F}_0(\alpha, s)$ is a concave function of $\alpha$ and
$$\left.\frac{\partial\tilde{F}_0(\alpha, s)}{\partial\alpha}\right|_{\alpha=0} = H(X_a|X_e, S = s)$$

With these properties we can find the conditions on $R_{\mathrm{SK}}$ and $R_{\mathrm{M}}$ for which $(R_{\mathrm{SK}}, E_{\mathrm{R}}, E_{\mathrm{S}})$ is strongly achievable. Taking derivative of (16) with respect to $\rho$, we can find that

$$R_{\mathrm{M}} = \frac{\partial E_0(\rho, p_S)}{\partial\rho}$$

when the optimal $\rho$ is in $(0, 1)$. For $R_{\mathrm{M}} \le \left.\frac{\partial E_0(\rho, p_S)}{\partial\rho}\right|_{\rho=0}$, $E_{\mathrm{R}}$ is maximized by $\rho = 0$ which corresponds to $E_{\mathrm{R}}(R_{\mathrm{M}}) = 0$. Thus $E_{\mathrm{R}}(R_{\mathrm{M}})$ is positive if

$$R_{\mathrm{M}} > \left.\frac{\partial E_0(\rho, p_S)}{\partial\rho}\right|_{\rho=0} = H(X_a|X_b, S) \tag{24}$$

by Theorem 3. Similarly, the secrecy exponent $E_{\mathrm{S}}(R_{\mathrm{M}}, R_{\mathrm{SK}})$ is positive if

$$R_{\mathrm{M}} + R_{\mathrm{SK}} < \left.\frac{\partial F_0(\alpha, p_S)}{\partial\alpha}\right|_{\alpha=0} = H(X_a|X_e, S). \tag{25}$$

In order to achieve positive $E_{\mathrm{R}}$ and $E_{\mathrm{S}}$, the key rate $R_{\mathrm{SK}}$ has to satisfy both (24) and (25). Namely,

$$\begin{aligned} R_{\mathrm{SK}} &< H(X_a|X_e, S) - H(X_a|X_b, S) \\ &= I(X_a; X_b|S) - I(X_a; X_e|S) \ . \end{aligned} \tag{26}$$

The upper bound (26) is the secret key capacity with degraded eavesdropper (13) if $p_S$ is the capacity achieving input distribution.

*Remark 1:* The results of Theorems 2 and 3, while derived for degraded channels, can also be applied to the non-degraded case. However, the used achievable binning scheme may not be capacity achieving in those settings. This applicability of these results follows from two observations. The first is that $\Pr(K_a \neq K_b)$ and $I(K_a; X_e^n, S^n, \Phi)$ respectively depends only on the marginal joint distributions $p_{X_a,X_b|S}(x_a, x_b|s)$ and $p_{X_a,X_e|S}(x_a, x_e|s)$. The second observation is that the reliability and secrecy exponents derived in the theorems are expressed in terms of $p_{X_a,X_b|S}(x_a, x_b|s)$ and $p_{X_a,X_e|S}(x_a, x_e|s)$, respectively.

Figure 1 depicts the achievable $(R_{\mathrm{SK}}, E_{\mathrm{R}}, E_{\mathrm{S}})$ for a fixed $p_S$. We can see that when $R_{\mathrm{SK}}$ is fixed, there is a tradeoff between $E_{\mathrm{R}}$ and $E_{\mathrm{S}}$ (see Section IV-A). Note that although we have control on $p_S \in \mathcal{P}_S$, the optimal $p_S$ that maximizes the reliability exponent and that maximizes secrecy exponent are in general different. For some case that Eve's observation is useless for estimating the key (e.g., the rich multipath channel model in II-C), we are interested in the reliability exponent. To find the largest $E_{\rm R}$, we express reliability exponent in terms of secret key rate $R_{\rm SK}$ by substituting $R_{\rm M} = H(X_a|X_e,S) - R_{\rm SK} - \varepsilon$ in (16). When $\varepsilon \to 0$, it is equal to operating at one extreme case of the tradeoff, i.e., $E_{\rm S} \to 0$, and yields a curve on the $R_{\rm SK}$-$E_{\rm R}$ plane in Figure 1. We optimize the reliability exponent over distribution $p_S \in \mathcal{P}_S$, namely,

$$E_{\rm R}(R_{\rm SK}) = \max_{p_S\in\mathcal{P}_S}\max_{0\le\rho\le1}\rho\left[H(X_a|X_e,S) - R_{\rm SK}\right] - E_0(\rho,p_S)\ . \tag{27}$$

### C. Gaussian EDMS

By Corollary 1 and for the Gaussian EDMS specified in Section II-C, the secret key capacity

$$C_{\rm K} = \max_{p_S\in\mathcal{P}_S} I(X_a;X_b|S)\ , \tag{28}$$

since $X_e$ is independent of $X_a$. $\mathcal{P}_S = \{p_S : E[|S|^2] \le \mathsf{P}\}$ is a set of distribution satisfying the power constraint. If a (not necessarily optimal) constant excitation signal $S_i = s$ where $|s|^2 = \mathsf{P}$ uniformly for all $1 \le i \le n$ is used, the resulting mutual information $I(X_a;X_b|S = s)$ is easily calculated to be

$$I_{\rm K}(\gamma_a,\gamma_b) = \log(1+\gamma_{\rm eq}), \tag{29}$$

$$\text{where} \qquad \gamma_{\rm eq} = \left(\frac{1}{\gamma_a}+\frac{1}{\gamma_b}+\frac{1}{\gamma_a\cdot\gamma_b}\right)^{-1}\ . \tag{30}$$

If, e.g., we fix $\gamma_a$ and let $\gamma_b$ grow without bound, the *equivalent* SNR $\gamma_{\rm eq}$ of the overall system is dominated by the worst SNR $\gamma_{\rm eq} \to \gamma_a$.

In the special case where Alice and Bob have the same SNR, i.e., $\gamma_a = \gamma_b = \gamma$, we denote

$$I_{\rm K}(\gamma) = \log(1+\gamma_{\rm eq}), \tag{31}$$

$$\text{where} \qquad \gamma_{\rm eq} = \frac{\gamma^2}{1+2\gamma}. \tag{32}$$

The following theorem specifies the secret key capacity of an equal-SNR Gaussian EDMS.

*Theorem 4:* The secret key capacity of Gaussian EDMS under power constraint $E[|S|^2] \le \gamma\sigma^2$ is

$$C_{\rm K}(\gamma) = \max_{0\le\lambda\le1}\lambda I_{\rm K}\left(\frac{\gamma}{\lambda}\right). \tag{33}$$

As we will discuss in Section IV-C, the optimal $\lambda$ is

$$\lambda_{\rm c} = \min\left\{\frac{\gamma}{\gamma_{\rm c}},1\right\}\ , \tag{34}$$

where $\gamma_{\rm c}$ is the positive root of the equation

$$I_{\rm K}(\gamma_{\rm c}) = \gamma_{\rm c}\cdot\frac{d}{d\gamma}I_{\rm K}(\gamma)\bigg|_{\gamma=\gamma_{\rm c}}. \tag{35}$$

Furthermore, an *on-off signal* which uses fraction $\lambda_{\rm c}$ of the channel can achieve the capacity.

With the above results we can characterize the minimum energy per key bit.

*Theorem 5:* The minimum energy per key bit of Gaussian EDMS is

$$\left(\frac{\mathcal{E}_{\rm b}}{\sigma^2}\right)_{\rm min} = \inf_{\gamma>0}\frac{\gamma}{C_{\rm K}(\gamma)}\log 2 = \frac{\gamma_{\rm c}}{I_{\rm K}(\gamma_{\rm c})}\log 2\ . \tag{36}$$

Our final results bound the Gaussian reliability exponent in terms of $R_{\rm SK}$ and SNR (cf. (27)) when the input is a constant excitation signal.

*Theorem 6:* If we define $I_{\rm c}(\gamma) = \log\left(\frac{1+\gamma_{\rm eq}}{2}\right)$ and $\rho^* = \exp(I_{\rm K}(\gamma) - R_{\rm SK}) - 1$, where $I_{\rm K}(\gamma)$ is defined in (31), then the reliability exponent given a uniform excitation signal, as a function of secret key rate and SNR breaks into three regions:

**Region 1** (high rate): If $R_{\rm SK} \ge I_{\rm K}(\gamma)$ then

$$E_{\rm R}(R_{\rm SK},\gamma) = 0.$$

**Region 2** (medium rate): If $I_{\rm K}(\gamma) > R_{\rm SK} \ge I_{\rm c}(\gamma)$ then

$$E_{\rm R}(R_{\rm SK},\gamma) = \rho^*[I_{\rm K}(\gamma) - R_{\rm SK} + 1] - (1+\rho^*)\log(1+\rho^*).$$

**Region 3** (low rate): If $I_{\rm c}(\gamma) > R_{\rm SK} \ge 0$ then

$$E_{\rm R}(R_{\rm SK},\gamma) = \begin{cases} I_{\rm K}(\gamma) - R_{\rm SK} + 1 - 2\log 2, & \text{if } \ \gamma_{\rm eq} \ge 1,\\ \rho^*[I_{\rm K}(\gamma) - R_{\rm SK} + 1] - (1+\rho^*)\log(1+\rho^*), & \text{if } \ 1 > \gamma_{\rm eq} \ge 0. \end{cases}$$

Although the optimal excitation signal to maximize Gaussian reliability exponent remains unknown, on-off signaling again improves the achieved exponent at low SNR. This improvement is discussed in Section IV.

## IV. Discussions

In this section we discuss various implications of the main results presented in Section III.

### A. Reliability-Secrecy Tradeoff

The problem of secret key generation from a correlated source can be formulated as a distributed source coding problem with a secrecy constraint. On the one hand, we want to understand the system reliability, which corresponds to the error exponent of a Slepian-Wolf problem [24]. On the other hand, we want to develop an analogous measure of secrecy, which corresponds to the privacy amplification considered in [25, 27]. A main consequence of Theorem 2 is that reliability and the secrecy can both be quantified for system and we observe that there is a fundamental tradeoff between the two.

From (24) and (25), we know that $E_{\rm R}$ is positive if $R_{\rm M} > H(X_a|X_b,S)$ and $E_{\rm S}$ is positive if $R_{\rm M} + R_{\rm SK} < H(X_a|X_e,S)$. More specifically, from (16) and (19), the public message rate $R_{\rm M}$ is a design parameter which allows us tradeoff $E_{\rm R}$ for $E_{\rm S}$. Figure 1 is a three dimensional plot of achievable $(R_{\rm SK},E_{\rm R},E_{\rm S})$ for a binary source from quantizing channel phase. Details are in Section IV-F. Figure 2(a) plots $E_{\rm R}$ vs $E_{\rm S}$ as a function of $R_{\rm M}$ for different values of $R_{\rm SK}$. By varying $R_{\rm M}$ for a fixed $R_{\rm SK}$, we get the $E_{\rm R}$-$E_{\rm S}$ tradeoff curve shown in Figure 2(b). This curve corresponds to a slice of the surface in Figure 1 at a constant $R_{\rm SK}$.

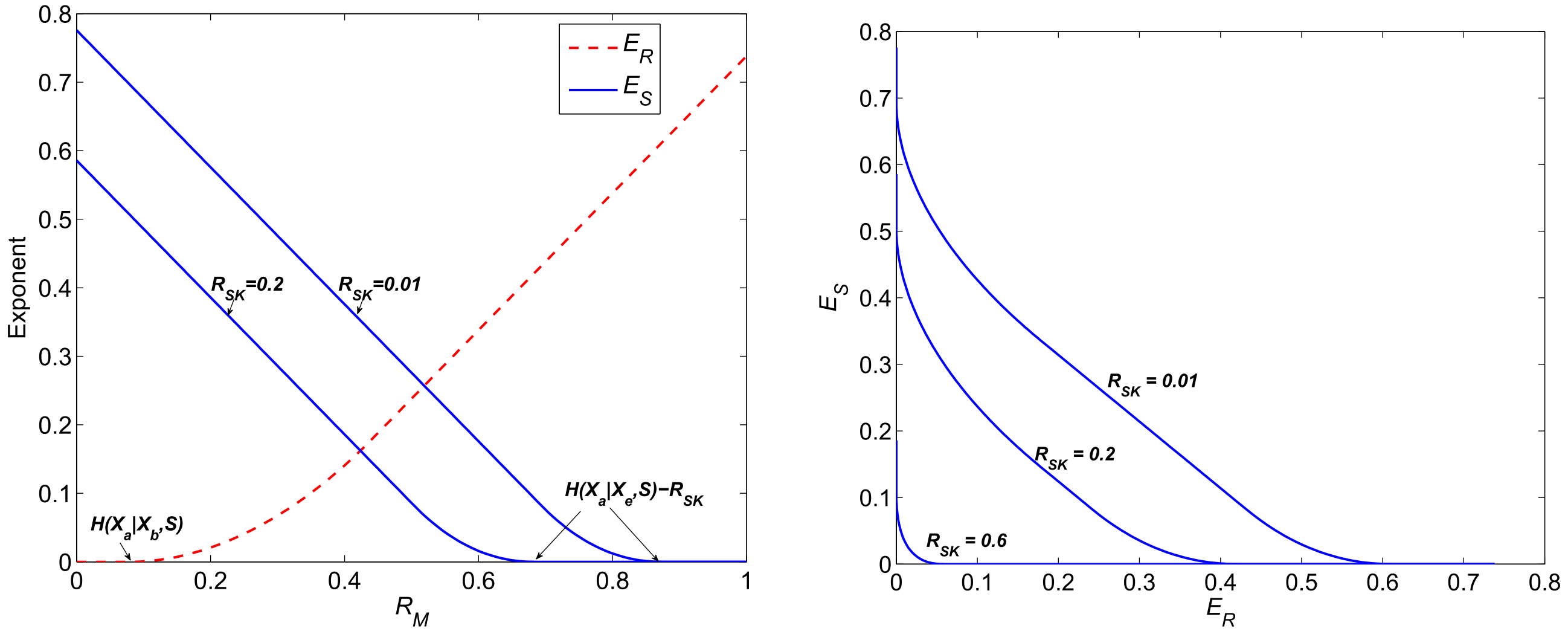

(a) (b)

Fig. 2. Binary source with $H(X_a|X_b,S)=0.08$ and $H(X_a|X_e,S)=0.88$. (a) $E_{\rm R}$ ($E_{\rm S}$) v.s $R_{\rm M}$ with $R_{\rm SK}=0.2$ and $R_{\rm SK}=0.01$ (b) $E_{\rm S}$-$E_{\rm S}$ tradeoff

### B. An Equivalent Noise Model of Gaussian EDMS

The secret key rate function (31) can be observed to have the same form as the AWGN channel capacity if we replace SNR $\gamma$ with $\gamma_{\rm eq}$. However, (31) is always less than AWGN channel capacity. To see this, note that from signal model (8), when $S$ is a constant excitation signal $s$ and $|s|^2=\mathsf{P}$, $X_a, X_b$ and $H$ form a Markov chain $X_a \leftrightarrow sH \leftrightarrow X_b$. Thus,

$$I(X_a;X_b)=I(sH;X_b)-I(sH;X_b|X_a)\ .$$

The first term has the same value as AWGN channel capacity while the second term is always non-negative. To obtain an insight of the secret key rate function $I_{\rm K}(\gamma)$, we present an *equivalent Gaussian noise model* in which the Bob's channel output $X_b$ is expressed as a combination of the component statistically aligned with $X_a$ and the component orthogonal to $X_a$, that is

$$X_b=\beta X_a+Z\ , \tag{37}$$

where

$$\beta=\frac{E[X_bX_a^*]}{E[|X_a|^2]}=\frac{\mathsf{P}}{\mathsf{P}+\sigma_a^2}=\frac{\gamma_a}{\gamma_a+1} \tag{38}$$

$$\begin{aligned} E[|Z|^2] &= E[|X_b|^2]-E[|\beta X_a|^2] \\ &= (\mathsf{P}+\sigma_b^2)-\frac{\mathsf{P}^2}{\mathsf{P}+\sigma_a^2}\ . \end{aligned} \tag{39}$$

This can also be seen as follows: $\beta X_a$ is the minimum mean-squared error (MMSE) estimate of $X_b$. Since $X_a$ and $X_b$ are jointly Gaussian with zero mean (recall that $S$ is known), the orthogonal error, $Z$, is also zero-mean Gaussian and independent of $X_a$ due to the orthogonality principle and Gaussianity. $X_a$ in (37) has the same distribution as in (8a) and $(X_a,X_b)$ has the same joint distribution as in (8). Figure 3 shows the relationship between $X_a$ and $X_b$ in the original and equivalent Gaussian model. The signal multiplication gain $\beta$ ($\beta<1$) is a deterministic function of Alice's SNR $\gamma_a$. The

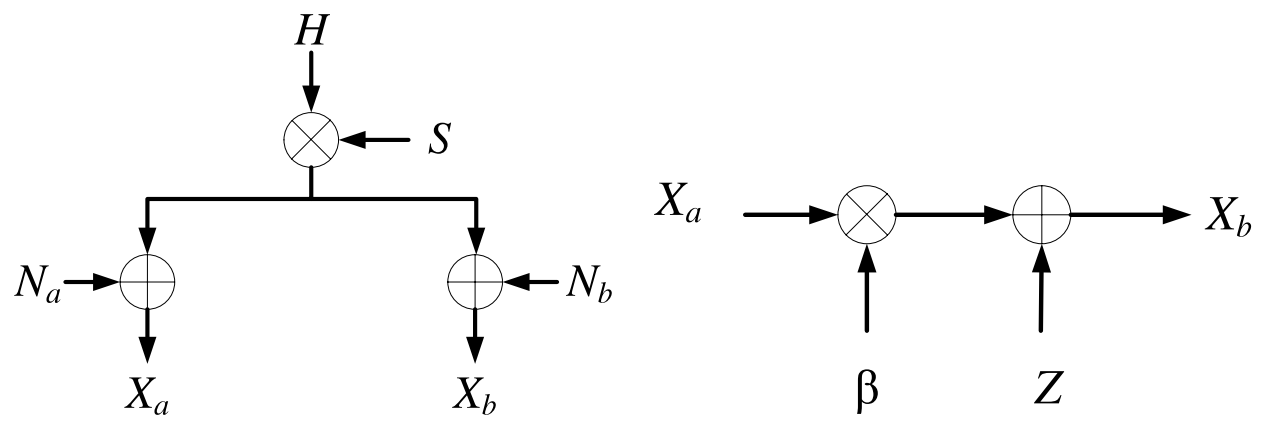

(a) (b)

Fig. 3. (a) Original signal model (b) Equivalent Gaussian noise model

signal to noise ratio of the equivalent model (37) is

$$\begin{aligned} \frac{E[|\beta X_a|^2]}{E[|Z|^2]} &= \frac{\frac{\mathsf{P}^2}{\mathsf{P}+\sigma_a^2}}{(\mathsf{P}+\sigma_b^2)-\frac{\mathsf{P}^2}{\mathsf{P}+\sigma_a^2}} \\ &= \left(\frac{\sigma_a^2+\sigma_b^2}{\mathsf{P}}+\frac{\sigma_a^2\sigma_b^2}{\mathsf{P}^2}\right)^{-1}=\gamma_{\rm eq}\ . \end{aligned}$$

This is the same result we get from calculating the mutual information given in ((30). For the symmetric SNR case, it reduces to (32). In the high-SNR regime, $\gamma_{\rm eq}\approx\gamma/2$ because $E[|X_a|^2]\approx\mathsf{P}$ and the end-to-end channel between $X_a$ and $X_b$ embeds two channel noises. The equivalent (compound) channel has multiplication gain $\beta\approx 1$ and the resulting secret key rate $I_{\rm K}(\gamma)\approx\log\left(1+\frac{\gamma}{2}\right)$ is a concave function.

On the other hand, in the low-SNR regime $\gamma_{\rm eq}\approx\gamma^2$ and $I_{\rm K}(\gamma)\approx\log(1+\gamma^2)\approx\gamma^2$ which is a convex function. Using the equivalent Gaussian noise model (37) when $\gamma\ll 1$, the signal multiplication gain $\beta\approx\gamma$ and

$$X_b=\beta X_a+Z\approx\gamma X_a+Z\ . \tag{40}$$

where $\frac{E[|X_a|^2]}{E[|Z|^2]}\approx 1$. This gain $\beta$ is linearly proportional to the operating SNR and the resulting secret key rate is $I_{\rm K}(\gamma)\approx\log\left(1+\gamma^2\right)\approx\gamma^2$, a convex function.

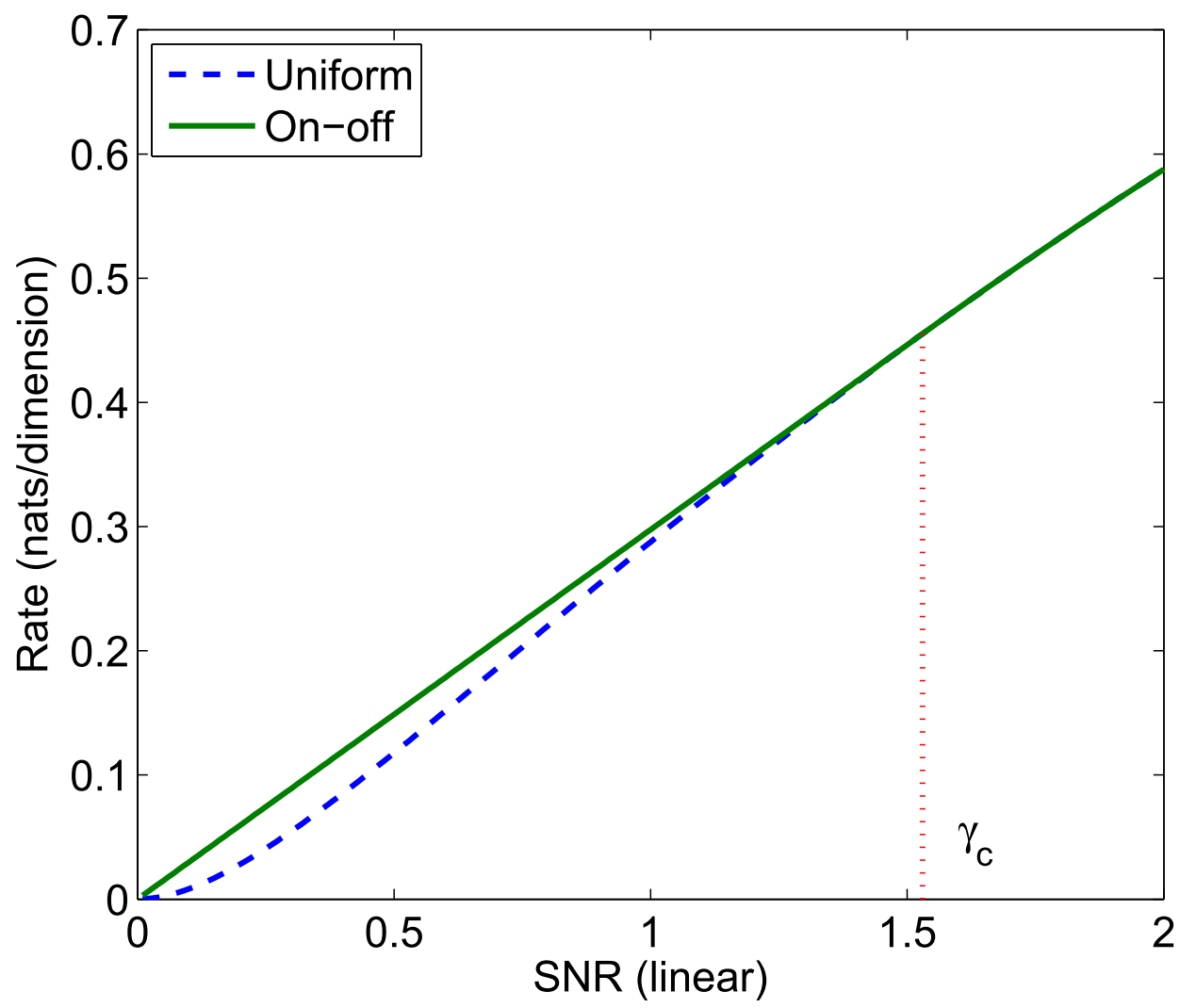

Fig. 4. Secret key rate in low SNR regime

### C. On-off Signaling Achieves Capacity

Due to the convexity of $I_{\rm K}(\gamma)$ in low SNR, an *on-off* type excitation signal with distribution

$$p_S(s) = (1-\lambda)\delta(s) + \lambda\delta(s-s_1)$$

with $s_1 = \sqrt{\gamma\sigma^2/\lambda}$ and $0 \le \lambda \le 1$, will achieve a higher secret key rate. The corresponding secret key rate is

$$\lambda I_{\rm K}\left(\frac{\gamma}{\lambda}\right) \ . \tag{41}$$

In the proof of Theorem 4 in Section V, we show that the secret key capacity is achieved by choosing $\lambda$ which maximizes (41). Taking the derivative of (41) with respect to $\lambda$, we can find the optimal parameter $\lambda$:

$$\lambda_{\rm c} \triangleq \arg\max_{0\le\lambda\le 1} \lambda I_{\rm K}\left(\frac{\gamma}{\lambda}\right) = \min\left\{\frac{\gamma}{\gamma_{\rm c}}, 1\right\} \tag{42}$$

where $\gamma_{\rm c}$ is the positive root of the equation

$$I_{\rm K}(\gamma_{\rm c}) = \gamma_{\rm c} \cdot \frac{d}{d\gamma} I_{\rm K}(\gamma)\bigg|_{\gamma=\gamma_{\rm c}} .$$

Solving $\gamma_{\rm c}$ numerically we find $\gamma_{\rm c} \approx 1.535$ (1.86 dB). Figure 4 shows the secret key rate for a uniform constant signal and an on-off signal.

The operational interpretation of this result is as follows. When $\gamma$ is below $\gamma_{\rm c}$, capacity is achieved by using the source $\lambda_{\rm c} n$ times, each at SNR $\gamma_{\rm c}$. The achievable key rate is equal to time-sharing between two SNRs (0 and $\gamma_{\rm c}$). The excitation signal thus becomes sparse and peaky when $\gamma$ is low. We note that an analogous result from an non-coherent communication in which a sparse/peaky signal also achieves channel capacity in low SNR [28, 29].

We can also interpret the improvement of the optimal on-off signal using the equivalent noise model developed in Section IV-B. From (40), in the low-SNR regime (in particular $\gamma \le \gamma_{\rm c}$), the multiplication gain $\beta$ of the equivalent channel approximates the operating SNR. When we use the optimal on-off signal, it results in a higher SNR ($\gamma_{\rm c}$). The resulting secret key rate is

$$\lambda_{\rm c} I_{\rm K}\left(\frac{\gamma}{\lambda_{\rm c}}\right) = \lambda_{\rm c} I_{\rm K}(\gamma_{\rm c}) \approx \underbrace{\lambda_{\rm c}\gamma_{\rm c}}_{\gamma}\gamma_{\rm c} \ .$$

The last approximation comes from $I_{\rm K}(\gamma) \approx \gamma^2$. Thus, the secret key rate-cost function approximates a linear function of $\gamma$ instead of a quadratic function.

One of the most significant effects due to the linearity of $C_{\rm K}(\gamma)$ in the low-SNR regime (because of the time-sharing effect) is the minimum energy per key bit $\left(\frac{\mathcal{E}_{\rm b}}{\sigma^2}\right)_{\rm min}$ given in Theorem 5. In particular, in contrast to the AWGN channel, $\left(\frac{\mathcal{E}_{\rm b}}{\sigma^2}\right)_{\rm min}$ is achieved in all $\gamma < \gamma_{\rm c}$. Figure 5(a) shows $\gamma/I_{\rm K}(\gamma)$ and $\gamma/C_{\rm K}(\gamma)$ as function of $\gamma$. Without using an on-off signaling, $\left(\frac{\mathcal{E}_{\rm b}}{\sigma^2}\right)_{\rm min}$ is achieved only at $\gamma_{\rm c}$.

### D. Gaussian Reliability Exponent Using An On-off Signal

In Theorem 6 we give an expression of reliability exponent for the Gaussian EDMS when the input is a constant excitation signal. Furthermore, from the discussion in Section IV-C, we can expect that an on-off signal can achieve higher reliability exponent for a given average SNR $\gamma$. Consider an excitation scheme that uses fraction $\lambda$ of the available channel degrees of freedom. To achieve a target $R_{\rm SK}$ with average SNR $\gamma$, the system operates at key rate $R_{\rm SK}/\lambda$ and SNR $\gamma/\lambda$. The error probability is upper bounded by $e^{-\lambda n E_{\rm R}(R_{\rm SK}/\lambda, \gamma/\lambda)}$ which has effective exponent

$$\lambda E_{\rm R}\left(\frac{R_{\rm SK}}{\lambda}, \frac{\gamma}{\lambda}\right) \ .$$

We maximize it over $0 \le \lambda \le 1$ to get *the error exponent optimal on-off signal*

$$\lambda_{\rm e} \triangleq \arg\max_{0\le\lambda\le 1} \lambda E_{\rm R}\left(\frac{R_{\rm SK}}{\lambda}, \frac{\gamma}{\lambda}\right) \tag{43}$$

and denote the error exponent achieved by the optimal on-off signal as

$$\bar{E}_{\rm R}(R_{\rm SK}, \gamma) = \lambda_{\rm e} E_{\rm R}\left(\frac{R_{\rm SK}}{\lambda_{\rm e}}, \frac{\gamma}{\lambda_{\rm e}}\right) \ . \tag{44}$$

Figure 6(a) and 6(b) show $E_{\rm R}(R_{\rm SK}, \gamma)$ and $\bar{E}_{\rm R}(R_{\rm SK}, \gamma)$ in low SNR as functions of $R_{\rm SK}$ and $\gamma$, respectively. Although the optimal input distribution for the error exponent is still unknown, we can see the results in Figure 6 that the on-off signal has a large improvement at low rates and low SNRs.

### E. Minimum Key Energy with A Finite Block Length

We use the developed reliability exponent to characterize minimum key energy with a finite block length code $\left(\frac{\mathcal{E}_{\rm key}}{\sigma^2}\right)_{\rm min}$. In contrast to minimum energy per key bit, $\left(\frac{\mathcal{E}_{\rm key}}{\sigma^2}\right)_{\rm min}$ is the minimum energy required to generate a $b_{\rm key}$-bit key by using a finite block length such that the system meets the reliability requirement $\Pr(K_a \neq K_b) \le \epsilon$. We provide an upper bound of such energy.

Consider using a source $n$ times and generate a $b_{\rm key}$-bit key. The secret key rate is $R_{\rm SK} = \log 2 \frac{b_{\rm key}}{n}$ and the energy

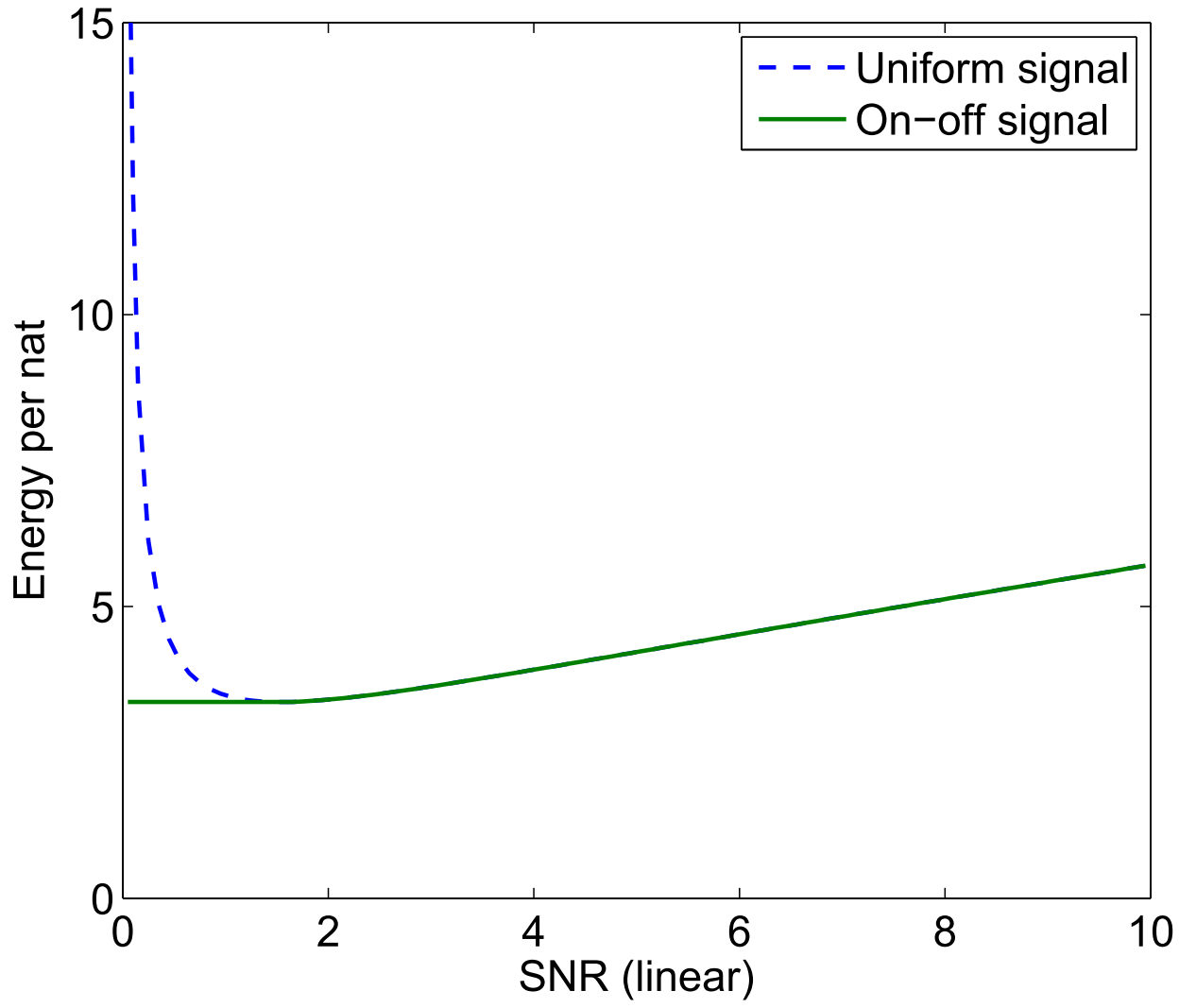

(a) Energy per key bit

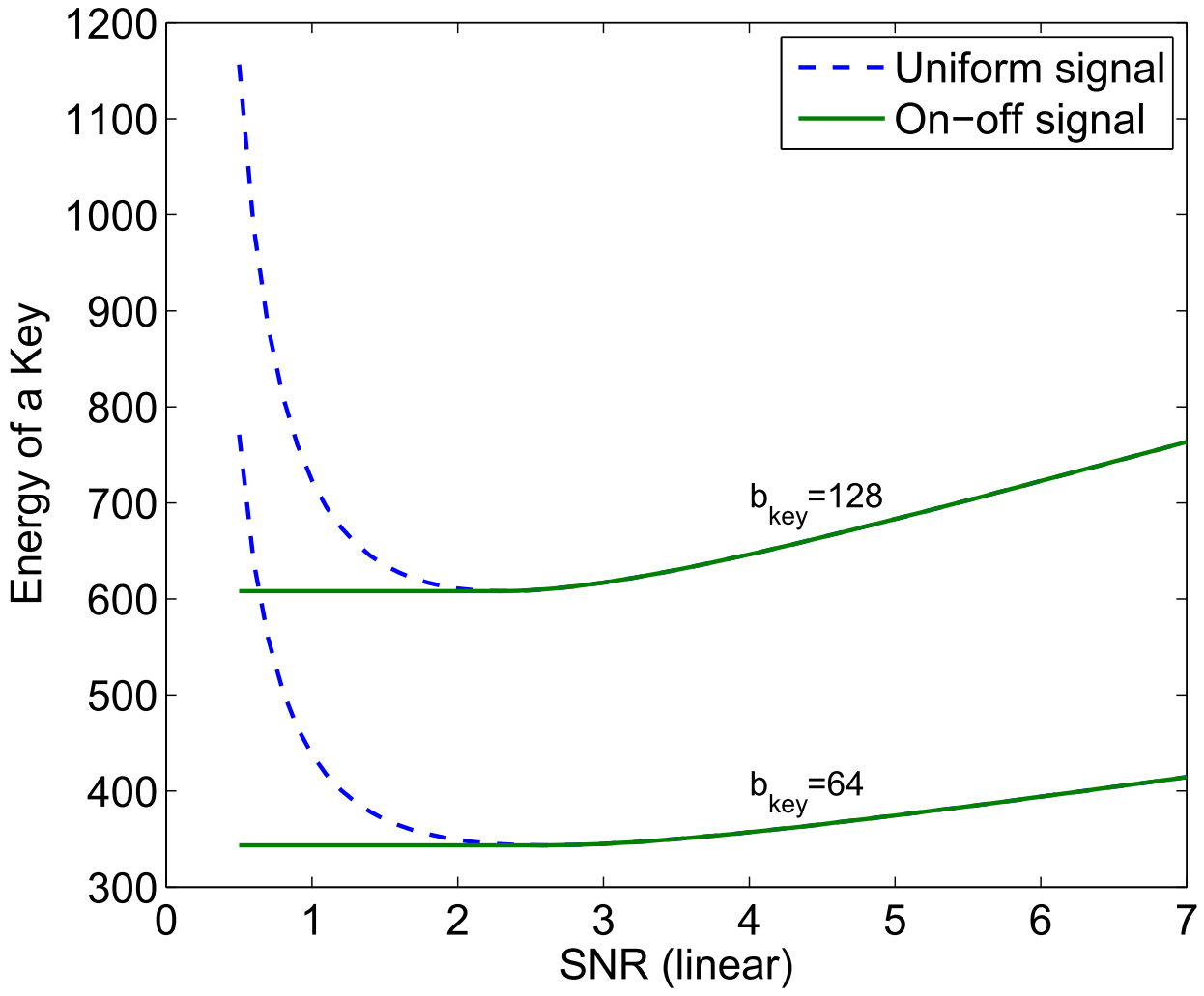

(b) Energy per key for $b_{\rm key} = 64$ and 128 bits. $\epsilon = 0.01$

Fig. 5. Energy of key generation

required to generate the key is $\mathcal{E}_{\rm key} = n\mathsf{P} = n\gamma\sigma^2$. Since $\mathcal{E}_{\rm key}$ is linearly proportional to $n$, we apply exponential bound of error probability we have developed to find the minimum block length $n$ needed to satisfy reliability condition. Namely,

$$J(n,\gamma,b_{\rm key}) \triangleq nE_{\rm R}\left(\log 2\frac{b_{\rm key}}{n},\gamma\right) \geq \log\frac{1}{\epsilon} \ . \quad (45)$$

Note that $J(n,\gamma,b_{\rm key})$ is an increasing function of $n$ and $\gamma$ when $b_{\rm key}$ is fixed. The normalized minimum energy of a $b_{\rm key}$-bit key can be bounded above as

$$\left(\frac{\mathcal{E}_{\rm key}}{\sigma^2}\right)_{\min} \leq \min_{p_S\in\mathcal{P}_S}\ \min_{\gamma\geq 0, n\in\mathbb{N}} \gamma n \quad (46)$$

subject to

$$J(n,\gamma,b_{\rm key}) \geq \log\frac{1}{\epsilon} \ ,$$

where $\mathcal{P}_S = \{p_S : E[|S|^2]/\sigma^2 \leq \gamma\}$. Given a $p_S \in \mathcal{P}_S$, (46) is a nonlinear optimization problem with a nonlinear constraint. We examine the upper bound by using a uniform excitation input and an optimal on-off excitation input defined in (43), the latter is equivalent to replacing $E_{\rm R}(\cdot,\cdot)$ with $\bar{E}_{\rm R}(\cdot,\cdot)$ given in (44). Figure 5(b) shows the numerical result of this upper bound for both $b_{\rm key} = 64$ and 128 as a function of $\gamma$ wherein $n$ is chosen to be the minimum integer satisfying the reliability constraint (45). We see the similar behavior as in minimum energy per bit: an optimal on-off excitation signal achieves the minimum value for all SNRs below a threshold while a constant excitation achieves this value only at threshold SNR.

### *F. An Example: Doubly Symmetric Binary Source from Quantized Channel Phases*

In a practical implementation of a key generation from reciprocal multipath channel, each user has to quantize its observation. We consider a simple (symmetric) phase quantization scheme [15]. In this scheme, Alice and Bob perform the same phase quantization that maps the random phase to binary level $\{0,1\}$, denoted by $Y_a$ and $Y_b$ respectively. The relation between $Y_a$ and $Y_b$ are characterized by a doubly symmetric binary source (DSBS($\theta$)) with transition probability $\theta = \Pr(Y_a \neq Y_b)$. In Appendix B, it is shown that with a constant excitation signal ($S = s$) input,

$$\theta = \frac{1}{2} - \frac{1}{\pi}\arctan\left(\sqrt{\gamma_{\rm eq}}\right) \ , \quad (47)$$

where $\gamma_{\rm eq}$ is defined in (32). Eve's quantized output $Y_e$ is assumed to be independent of $Y_a$ and $Y_b$ in a rich multipath case.

*1) Secret key capacity:* The achievable secret key rate with a constant excitation signal is

$$I_{\rm K}(\gamma) = I(Y_a;Y_b|S) = 1 - H_B(\theta) \ , \quad (48)$$

where $H_B(\theta) = -\theta\log\theta - (1-\theta)\log(1-\theta)$ is the binary entropy function. Similar to a Gaussian source, the $I_{\rm K}(\gamma)$ again exhibits convexity in the low-SNR regime. The optimality of an on-off signal to achieve capacity follows via the same reason as Theorem 4 though the result $\gamma_{\rm c}$ is naturally different from Gaussian case. The optimal operating SNR for the binary case is $\gamma_{\rm c} \approx 1.28$ (1.07 dB).

*2) Reliability exponent:* In Appendix C, we derive the reliability exponent of the this DSBS in terms of message rate $R_{\rm M}$ and $\theta$ (47) when a constant excitation signal is applied. Since Eve has independent channel output, the conditional entropy of quantized binary variable $H(Y_a|Y_e,S) = 1$. From (25), the key rate $R_{\rm SK}$ satisfying the secrecy condition is $R_{\rm M} = H(Y_a|S=s) - R_{\rm SK} = 1 - R_{\rm SK}$. Thus, we get reliability exponent in terms of $R_{\rm SK}$:

**Region 1** (high rate): If $R_{\rm SK} \geq I_{\rm K}(\gamma)$,

$$E_{\rm R}(R_{\rm SK},\gamma) = 0 \ .$$

where $I_{\rm K}(\gamma)$ is defined in (48).

**Region 2** (medium rate): If $I_{\rm K}(\gamma) > R_{\rm SK} \geq I_{\rm c}(\gamma)$ then

$$E_{\rm R}(R_{\rm SK},\gamma) = T_\theta(\tau) - H_B(\tau) \ ,$$

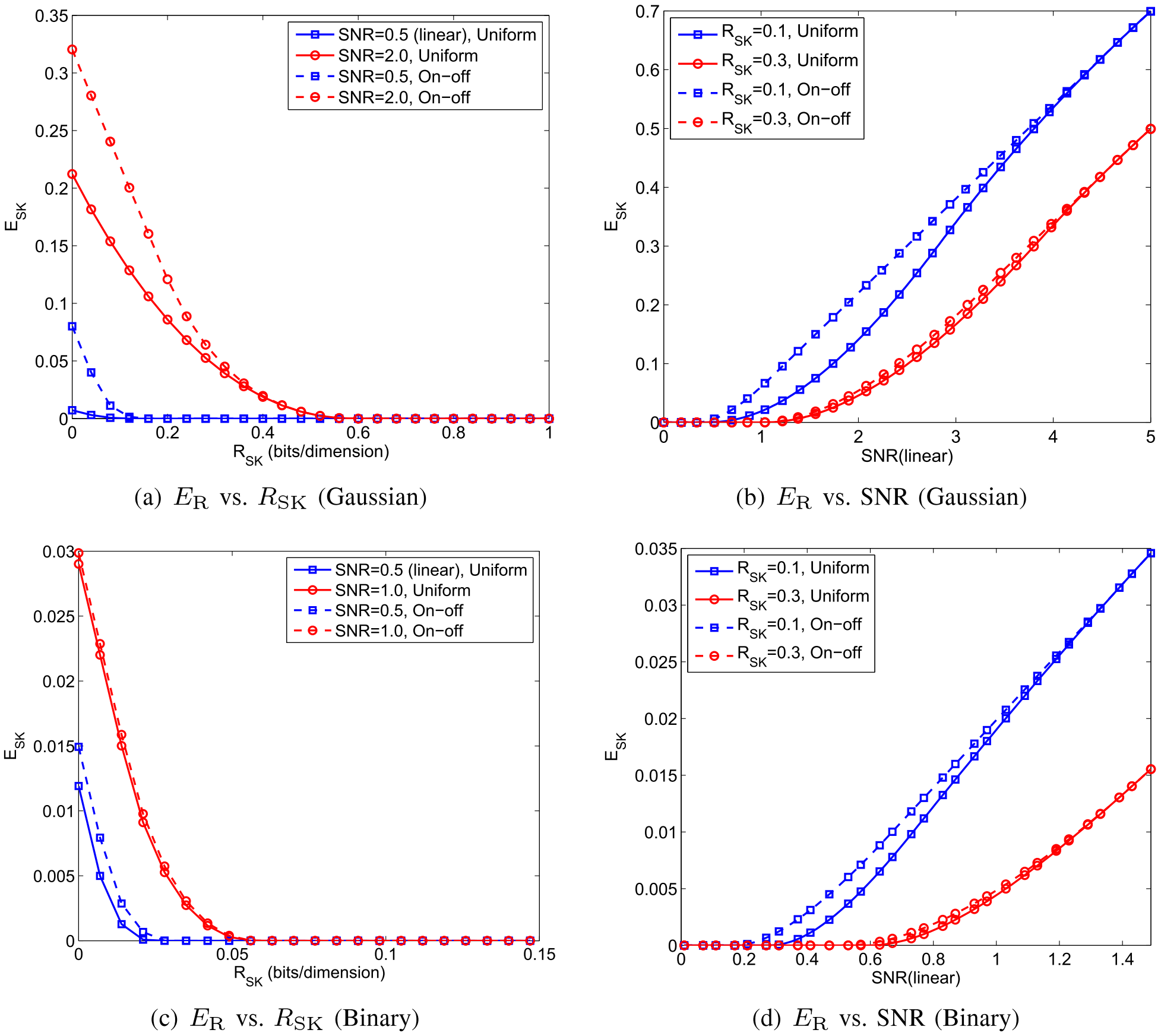

(a) $E_{\mathrm{R}}$ vs. $R_{\mathrm{SK}}$ (Gaussian)

(b) $E_{\mathrm{R}}$ vs. SNR (Gaussian)

(c) $E_{\mathrm{R}}$ vs. $R_{\mathrm{SK}}$ (Binary)

(d) $E_{\mathrm{R}}$ vs. SNR (Binary)

Fig. 6. Error exponent of low-SNR for Gaussian and quantized channels

where $\theta$ is defined in (47) and

$$I_c(\gamma) = 1 - H_B\left(\frac{\sqrt{\theta}}{\sqrt{\theta}+\sqrt{1-\theta}}\right) , \tag{49}$$

$$\tau = H_B^{-1}(1 - R_{\mathrm{SK}}) , \tag{50}$$

$$T_\theta(\tau) = -\tau\log(\theta) - (1-\tau)\log(1-\theta) . \tag{51}$$

**Region 3** (low rate): If $I_c > R_{\mathrm{SK}} \geq 0$ then

$$E_{\mathrm{R}}(R_{\mathrm{SK}}, \gamma) = 1 - 2\log(\sqrt{\theta} + \sqrt{1-\theta}) - R_{\mathrm{SK}} .$$

Figure 6(c) and 6(d) show the error exponent in the low SNR regime with constant and on-off input signals. Similar to the Gaussian case, the error exponent of an on-off signal improves for low SNR and for low rate.

*3) Secrecy exponent:* The result above consider a case that Eve's $Y_e$ is independent of $Y_a$ and $Y_b$. Regarding the secrecy condition, it is insightful to consider the correlated eavesdropping. Let $Y_e$ is correlated with $Y_a$ by a BSC with transition probability $w = \Pr(Y_a \neq Y_e) \leq 1/2$. From (19)-(21) and after some manipulation, we can find that

$$F_0(\alpha, p_S) = -\log\left(w^{1+\alpha} + (1-w)^{1+\alpha}\right)$$

$$\text{and} \quad \frac{\partial F_0(\alpha, p_S)}{\partial \alpha} = T_w(\delta) ,$$

$$\text{where} \quad \delta = \frac{w^{1+\alpha}}{w^{1+\alpha} + (1-w)^{1+\alpha}}$$

and $T_w(\delta)$ is defined the same as (51). The secrecy exponent is obtained by optimizing over $0 \leq \alpha \leq 1$ and can be expressed as

$$E_{\mathrm{S}}(R_{\mathrm{SK}}, R_{\mathrm{M}}) =$$
$$\begin{cases} 0, & R_{\mathrm{SK}} + R_{\mathrm{M}} > H_B(w) \\ F_0(\alpha^*, p_S) - \alpha^*(R_{\mathrm{SK}} + R_{\mathrm{M}}), & R_c < R_{\mathrm{SK}} + R_{\mathrm{M}} \leq H_B(w) \\ F_0(1, p_S) - (R_{\mathrm{SK}} + R_{\mathrm{M}}), & R_{\mathrm{SK}} + R_{\mathrm{M}} \leq R_c \end{cases}$$

where $\alpha^*$ is the solution of $T_w(\delta) = R_{\mathrm{SK}} + R_{\mathrm{M}}$ and

$$R_c = \left.\frac{\partial F_0(\alpha, p_S)}{\partial \alpha}\right|_{\alpha=1} .$$

Figure 2 shows the case where $\theta = 0.01$ and $w = 0.3$. This corresponds to $H(X_a|X_b, S) = 0.08$ and $H(X_a|X_e, S) = 0.881$.

## V. Proofs of Main Results

### A. Proof of Theorem 1

To simplify notation let $X_a^s, X_b^s, X_e^s$ denote a triplet of random variables with the same distribution as that induced on $X_a, X_b, X_e$ given $S = s$. In other words, $p_{X_a^s, X_b^s, X_e^s}(x_a, x_b, x_e) = p_{X_a, X_b, X_e|S}(x_a, x_b, x_e|s)$. The value of $s$ is the *state* of the system. The state-dependent secret key capacity $C_{\mathrm{K}}(s)$ follows immediately from [7, Theorem 1]:

$$C_{\mathrm{K}}(s) = \max_{U^s T^s} I(T^s; X_b^s|U^s) - I(T^s; X_e^s|U^s). \tag{52}$$

The capacity is found by maximizing over a pair of state-dependent random variables $U^s$ and $T^s$ where the overall distribution is $p_{U^s,T^s}(u,t)p_{X_a^s|T^s}(x_a|t)p_{X_b^s,X_e^s|X_a^s}(x_b,x_e|x_a)$.

Since the excitation signal $s^n$ is designer-chosen and need not be constant, i.e., we need not $s_i = s$ for all $i$ and some choice of $s$. For any choice of $s^n$ such that $P_{s^n} \to p_S \in \mathcal{P}_S$ as $n \to \infty$, we have an achievable secret key rate $\sum_s p_S(s)C_{\rm K}(s)$. Combine the state-dependent auxiliary random variables $U^s,T^s$ and the state distribution $p_S$, the auxiliary random variable $U,T$ has joint distribution given in (11). We next upper bound the achievable rate to show that maximizing $\sum_s p_S(s)C_{\rm K}(s)$ yields the secret key capacity of an EDMS.

For the converse part of Theorem 1, we want to show that for any $P_{s^n} \in \mathcal{P}_S$ and secret key rate $R_{\rm SK}$ satisfying (3)-(5), then there is some $p_S \in \mathcal{P}_S$ and $U,T$ such that

$$\begin{aligned} R_{\rm SK} &\le [I(T;X_b|U,S) - I(T;X_e|U,S)] \\ &= \sum_s p_S(s)\Big[I(T;X_b|U,S=s) - I(T;X_e|U,S=s)\Big] \ , \end{aligned}$$

cf. (10).

The proof is similar to the proof in [7] and use the following lemma [7, Lemma 4.1]

*Lemma 1:* For arbitrary random variables $U,V$ and sequences $X_b^n, X_e^n$, we have

$$\begin{aligned} &I(U;X_b^n|V) - I(U;X_e^n|V) \\ &= \sum_{i=1}^n \left[I(U;X_{b,i}|X_b^{i-1}X_{e,i+1}^n V) - I(U;X_{e,i}|X_b^{i-1}X_{e,i+1}^n V)\right] \end{aligned}$$

Proceeding to the proof of the converse, we show the following set of inequalities for any sequence $s^n$:

$$\begin{aligned} &nR_{\rm SK} \\ &\overset{\text{(i)}}{\le} H(K_a|S^n=s^n) + n\epsilon \\ &= I(K_a;X_b^n,\Phi|S^n=s^n) + H(K_a|X_b^n,\Phi,S^n=s^n) + n\epsilon \\ &\overset{\text{(ii)}}{\le} I(K_a;X_b^n,\Phi|S^n=s^n) + 2n\epsilon \\ &\overset{\text{(iii)}}{\le} I(K_a;X_b^n,\Phi|S^n=s^n) - I(K_a;X_e^n,\Phi|S^n=s^n) + 3n\epsilon \\ &= I(K_a;X_b^n|\Phi,S^n=s^n) - I(K_a;X_e^n|\Phi,S^n=s^n) + 3n\epsilon \\ &\overset{\text{(iv)}}{=} \sum_{i=1}^n \Big[I(K_a;X_{b,i}|X_b^{i-1}X_{e,i+1}^n\Phi S_{n\backslash i}=s_{n\backslash i},S_i=s_i) \\ &- I(K_a;X_{e,i}|X_b^{i-1}X_{e,i+1}^n\Phi S_{n\backslash i}=s_{n\backslash i},S_i=s_i)\Big] + 3n\epsilon \\ &\overset{\text{(v)}}{=} \sum_{i=1}^n I(K_a;X_{b,i}|U_i,S_i=s_i) \\ &\qquad - I(K_a;X_{e,i}|U_i,S_i=s_i) + 3n\epsilon \end{aligned} \tag{53}$$

where (i) follows from (3), (ii) from Fano's inequality and (iii) from the secrecy condition (4). The equality in (iv) follows from Lemma 1 where $S_{n\backslash i}$ denotes $S^{i-1}S_{i+1}^n$. Finally, (v) follows from letting $U_i = X_b^{i-1}X_{e,i+1}^n\Phi S_{n\backslash i}$. Letting $J$ be a random variable independent of all others and uniformly distributed over $\{1,2,\ldots,n\}$. The last sum can be written as

$$\begin{aligned} &n\sum_{i=1}^n \Pr(J=i)\Big[I(K_a;X_{b,i}|U_i,S_i=s_i,J=i) \\ &\qquad - I(K_a;X_{e,i}|U_i,S_i=s_i,J=i)\Big] \\ &= n\Big[I(K_a;X_{b,J}|U_J,S_J=s_J,J) \\ &\qquad - I(K_a;X_{e,J}|U_J,S_J=s_J,J)\Big] \\ &= n\sum_{s\in\mathcal{S}} P_{s^n}(s)\Big[I(K_a;X_{b,J}|U_J,J,S_J=s) \\ &\qquad - I(K_a;X_{e,J}|U_J,J,S_J=s)\Big] \\ &\le n\sum_{s\in\mathcal{S}} p_S(s)\Big[I(K_a;X_{b,J}|U,S_J=s) \\ &\qquad - I(K_a;X_{e,J}|U,S_J=s)\Big] \end{aligned}$$

for some $p_S \in \mathcal{P}_S$. The final inequality is due to the fact that type $P_{s^n}$ is a subset of distributions in $\mathcal{P}_S$ and $U=(U_J,J)$. Letting $T=(K_a,U)$, the construction of $U$, $T$, $X_{a,J}$, $X_{b,J}$, $X_{e,J}$ satisfies condition (11) for a given $S_J=s$ and ($X_{a,J}$, $X_{b,J}$, $X_{e,J}$, $S_J$) have the same joint distribution as ($X_a$, $X_b$, $X_e$, $S$). This demonstrates the tightness of (10).

We next develop the upper bound on the secret key rate specified in (12). The development of the bound agrees with the above development up to step (ii) in equation (53). From there we continue as below:

$$\begin{aligned} nR_{\rm SK} &< I(K_a;X_b^n,\Phi|S^n=s^n) + 2n\epsilon \\ &\le I(K_a;X_b^n,X_e^n,\Phi|S^n=s^n) + 2n\epsilon \\ &\overset{\text{(vi)}}{<} I(K_a;X_b^n|X_e^n,\Phi,S^n=s^n) + 3n\epsilon \\ &= I(K_a,\Phi;X_b^n|X_e^n,S^n=s^n) \\ &\qquad - I(\Phi;X_b^n|X_e^n,S^n=s^n) + 3n\epsilon \\ &\overset{\text{(vii)}}{\le} I(X_a^n;X_b^n|X_e^n,S^n=s^n) + 3n\epsilon \\ &\overset{\text{(viii)}}{=} n\sum_{s\in\mathcal{S}} P_{s^n}(s)I(X_a;X_b|X_e,S=s) + 3n\epsilon \\ &\le n\sum_{s\in\mathcal{S}} p_S(s)I(X_a;X_b|X_e,S=s) + 3n\epsilon \\ &= nI(X_a;X_b|X_e,S) + 3n\epsilon \end{aligned}$$

for some $p_S \in \mathcal{P}_S$. The inequality (vi) comes from the chain rule of mutual information and the secrecy condition (4). That in (vii) follows the fact that mutual information is non-negative and $K_a$ and $\Phi$ are function of $X_a^n$ and $S^n$, implying the Markov relation $\Phi,K_a \leftrightarrow S^n,X_a^n \leftrightarrow X_b^n,X_e^n$. Finally, (viii) uses the memoryless property of the source (cf. (1)) and is expressed in terms of the type $P_{s^n}$.

## *B. Proof of Corollary 1*

Given any particular state $S=s$ an EDMS is a DMS. Recall that a DMS with degraded states is one in which for every $s \in \mathcal{S}$ either $X_e$ is a degraded version of $X_b$ with respect to $X_a$ (i.e., $X_e^s \leftrightarrow X_b^s \leftrightarrow X_a^s$), or the reverse. Defining the set

$$\mathcal{D} = \{s \in \mathcal{S} : X_e \text{ is a degraded version of } X_b \text{ in state } s\} .$$

By the specific choice of $T^s = X_a^s$ for $s \in \mathcal{D}$, $T^s = \emptyset$ for $s \in \mathcal{D}^c$, and $U^s = \emptyset\ \forall s \in S$, we have from [7, Theorem 1] that

$$C_{\mathrm{K}}(s) = \begin{cases} I(X_a; X_b|S = s) - I(X_a; X_e|S = s) & \text{, if } s \in \mathcal{D} \\ 0 & \text{, else} \end{cases}$$

Average over all states and we arrive at

$$\begin{aligned} C_{\mathrm{K}} &= \max_{p_S \in \mathcal{P}_S} \sum_{s \in \mathcal{S}} p_S(s) C_{\mathrm{K}}(s) \\ &= \max_{p_S \in \mathcal{P}_S} \sum_{s \in \mathcal{S}} p_S(s) |I(X_a; X_b|S = s) - I(X_a; X_e|S = s)|^+, \end{aligned}$$

where the input constraint $p_S \in \mathcal{P}_S$ may force the system to put non-zero probability mass on reversely degraded states $s \in \mathcal{D}^c$. Finally, if the eavesdropper is degraded (in all states) we can drop the $|\cdot|^+$ and average the mutual information expression with respect to $p_S$ giving (13).

### C. Proof of Theorem 2

It is sufficient to prove the region $\tilde{\mathcal{R}}(p_S, R_{\mathrm{M}})$ satisfying (22) is achievable. To prove the achievability, we use an $(n, R_{\mathrm{SK}}, R_{\mathrm{M}})$ random binning secret key system according to Definition 11.

*1) Proof of Reliability Exponent:* We recall from that definition that the coding scheme involves two random binnings functions: the key binning $f_a(\cdot\,; s^n)$, cf. (14), and the public message binning $g(\cdot\,; s^n)$, cf. (15). All that remains to specify the system fully is to define Bob's decoding function $f_b(\cdot\,; s^n)$. Bob's decoding function will be a concatenation of a maximum likelihood decoder with random key binning function.

For compactness of notation let $W_{s^n}(x_a^n|x_b^n) = p_{X_a^n|X_b^n,S^n}(x_a^n|x_b^n, s^n)$. Bob's ML estimate of $X_a^n$ is

$$\hat{X}_a^n = h(\phi, x_b^n, s^n) = \underset{x_a^n : g(x_a^n; s^n) = \phi}{\arg\max}\ W_{s^n}(x_a^n|x_b^n)$$

and generates his key as

$$K_b = f_a(h(\Phi, X_b^n, S^n); s^n).$$

Define $\mathcal{E}_k$ to be the event that $K_a \neq K_b$ and $\mathcal{E}_x$ to be the event that $X_a^n \neq \hat{X}_a^n$. We bound the error probability by the probability of making an erroneous estimate of $X_a^n$.

$$\begin{aligned} \Pr(K_a \neq K_b) &= \Pr(\mathcal{E}_k \cap \mathcal{E}_x) + \Pr(\mathcal{E}_k \cap \mathcal{E}_x^c) \\ &= \Pr(\mathcal{E}_k \cap \mathcal{E}_x) \leq \Pr(\mathcal{E}_x). \end{aligned}$$

In [24] Gallager analyzes the *ensemble* random binning code and ML decoding for Slepian-Wolf coding and bounds $\Pr(\mathcal{E}_x)$ as

$$\Pr(\mathcal{E}_x) \leq |\mathcal{M}|^{-\rho} \sum_{x_b^n} Q_{s^n}(x_b^n) \left( \sum_{x_a^n} W_{s^n}(x_a^n|x_b^n)^{\frac{1}{1+\rho}} \right)^{1+\rho} \tag{54}$$

for all $0 \leq \rho \leq 1$. When the source and channel are memoryless, (54) simplifies further as,

$$\begin{aligned} &\Pr(K_a \neq K_b) \\ &\leq |\mathcal{M}|^{-\rho} \prod_{i=1}^{n} \sum_{x_b} Q_{s_i}(x_b) \left( \sum_{x_a} W_{s_i}(x_a|x_b)^{\frac{1}{1+\rho}} \right)^{1+\rho} \\ &= \exp\Bigg\{ -n \Bigg[ \rho R_{\mathrm{M}} \ldots \\ &- \sum_{s \in \mathcal{S}} P_S(s) \log \Big( \sum_{x_b} Q_s(x_b) \Big( \sum_{x_a} W_s(x_a|x_b)^{\frac{1}{1+\rho}} \Big)^{1+\rho} \Big) \Bigg] \Bigg\}. \end{aligned}$$

where we choose the type $P_{s^n}(s)$ of the excitation signal $s^n$ to converge to $p_S(s)$ as $n \to \infty$. Taking the logarithm on both sides and maximizing with respect to $\rho, 0 \leq \rho \leq 1$, we get (16).

*2) Proof of Secrecy Exponent:* To develop the secrecy exponent, we build off the "privacy amplification" analysis technique used in [25] with $\alpha = 1$ and in [26] for wire-tap channel problem. As discussed in Sec. III, in those papers $X_a = X_b$, i.e., Alice and Bob's observations are the same with probability one. Herein we incoporate the effect of the public message and of the excitation signal into the analysis technique.

The analysis uses the Renyi's entropy of order $1 + \alpha$ with $0 \leq \alpha \leq 1$. The Renyi entropy of order $1 + \alpha$ of a random variable $X$ with distribution function $p_X(\cdot)$ is

$$H_{1+\alpha}(X) = -\frac{1}{\alpha} \log \sum_x p_X(x)^{1+\alpha} .$$

We can interpret Renyi entropy in terms of an independent but identically distributed random variable $X'$, i.e., $p_{X,X'}(a, b) = p_X(a)p_X(b)$, as follows

$$\begin{aligned} H_{1+\alpha}(X) &= -\frac{1}{\alpha} \log \sum_x p_X(x) \left( \sum_a p_X(a) \mathbf{1}(x = a) \right)^{\alpha} \\ &= -\frac{1}{\alpha} \log \sum_x p_X(x) \Pr(X' = x)^{\alpha} \end{aligned}$$

where $\mathbf{1}(\cdot)$ is the indicator function.

One property of the Renyi entropy we find useful is that it is upper bounded by the Shannon entropy. It follows from Jensen's inequality:

$$\begin{aligned} H_{1+\alpha}(X) &= -\frac{1}{\alpha} \log \left( \sum_x p_X(x) p_X(x)^{\alpha} \right) \\ &\leq -\frac{1}{\alpha} \sum_x p_X(x) \log p_X(x)^{\alpha} \\ &= -\sum_x p_X(x) \log p_X(x) = H(X). \end{aligned}$$

Let $\mathcal{C}$ denote a specific code and let $\mathscr{C}$ denote a randomly chosen code from some ensemble. We begin the exponential characterization of secrecy by showing that, as blocklength gets large, the average mutual information leakage of the ensemble of random binning secret key codes can be bounded as

$$\liminf_{n\to\infty} -\frac{1}{n} E_{\mathscr{C}}[I(K_a; X_e^n, \Phi | S^n = s^n, \mathscr{C})] \geq E_{\mathrm{S}}(R_{\mathrm{M}}, R_{\mathrm{SK}}) + o(1).$$

Based on this result we will show that there exist a deterministic code that achieves the same secrecy exponent.

Rewrite the mutual information $I(K_a; X_e^n, \Phi | S^n = s^n, \mathscr{C} = \mathcal{C})$, for the specific code choice $\mathscr{C} = \mathcal{C}$, as

$$\begin{aligned} & I(K_a; X_e^n, \Phi | S^n = s^n, \mathscr{C} = \mathcal{C}) \\ & = H(K_a | S^n = s^n, \mathscr{C} = \mathcal{C}) - H(K_a | X_e^n, \Phi, S^n = s^n, \mathscr{C} = \mathcal{C}) \\ & = H(K_a | S^n = s^n, \mathscr{C} = \mathcal{C}) - H(K_a, \Phi | X_e^n, S^n = s^n, \mathscr{C} = \mathcal{C}) \\ & \quad + H(\Phi | X_e^n, S^n = s^n, \mathscr{C} = \mathcal{C}) \\ & \leq nR_{\mathrm{SK}} + nR_{\mathrm{M}} - H(K_a, \Phi | X_e^n, S^n = s^n, \mathscr{C} = \mathcal{C}) \ . \end{aligned} \tag{55}$$

We use the following notation for convenience

$$\begin{aligned} & \tilde{Q}_{s^n}(x_e^n) \equiv p_{X_e^n|S^n}(x_e^n|s^n), \\ & V_{s^n}(x_a^n|x_e^n) \equiv p_{X_a^n|X_e^n,S^n}(x_a^n|x_e^n, s^n) \\ & \tilde{V}_{s^n,\mathcal{C}}(k, \phi|x_e^n) \equiv p_{K_a,\Phi|X_e^n,S^n,\mathscr{C}}(k, \phi|x_e^n, s^n, \mathcal{C}) \qquad (56) \\ & \quad = \sum_{x_a^n \in \mathcal{X}_a^n} p_{X_a^n|X_e^n,S^n}(x_a^n|x_e^n, s^n) \mathbf{1}[k, \phi|x_a^n, s^n, \mathcal{C}] \\ & \quad = \sum_{x_a^n \in \mathcal{X}_a^n} V_{s^n}(x_a^n|x_e^n) \mathbf{1}[k, \phi|x_a^n, s^n, \mathcal{C}] \end{aligned}$$

where use use the shorthand notation $\mathbf{1}[k, \phi|x_a^n, s^n, \mathcal{C}]$, defined as

$$\mathbf{1}[k, \phi|x_a^n, s^n, \mathcal{C}] = \mathbf{1}[k = f_{a,\mathcal{C}}(x_a^n, s^n), \phi = g_{\mathcal{C}}(x_a^n, s^n)]. \tag{57}$$

The functions $f_{a,\mathcal{C}}(\cdot)$ and $g_{\mathcal{C}}(\cdot)$ are the binning functions of the specific codebook choice $\mathscr{C} = \mathcal{C}$ that is being conditioned on. In the following we lower bound the expectation over the choice of code of the conditional entropy term in (55).

$$\begin{aligned} & E_{\mathscr{C}}[H(K_a, \Phi | X_e^n, S^n = s^n, \mathscr{C})] \\ & = E_{\mathscr{C}}\left[\sum_{x_e^n} \tilde{Q}_{s^n}(x_e^n) H(K_a, \Phi | X_e^n = x_e^n, S^n = s^n, \mathscr{C})\right] \\ & \geq E_{\mathscr{C}}\left[\sum_{x_e^n} \tilde{Q}_{s^n}(x_e^n) H_{1+\alpha}(K_a, \Phi | X_e^n = x_e^n, S^n = s^n, \mathscr{C})\right] \\ & \overset{(a)}{=} E_{\mathscr{C}}\left[\sum_{x_e^n} \tilde{Q}_{s^n}(x_e^n) \frac{-1}{\alpha} \log\left(\sum_{k,\phi} \tilde{V}_{s^n,\mathscr{C}}(k, \phi|x_e^n) \times \right.\right. \\ & \qquad \left.\left. \Pr\Big((K', \Phi') = (k, \phi) | X_e^n = x_e^n, S^n = s^n, \mathscr{C}\Big)^{\alpha}\right)\right] \\ & \overset{(b)}{\geq} \sum_{x_e^n} \tilde{Q}_{s^n}(x_e^n) \frac{-1}{\alpha} \log\left(E_{\mathscr{C}}\left[\sum_{k,\phi} \tilde{V}_{s^n,\mathscr{C}}(k, \phi|x_e^n) \times \right.\right. \\ & \qquad \left.\left. \Pr\Big((K', \Phi') = (k, \phi) | X_e^n = x_e^n, S^n = s^n, \mathscr{C}\Big)^{\alpha}\right]\right) \end{aligned} \tag{58}$$

where in (a), $K'_a = f_{a,\mathscr{C}}(X'^n_a; s^n)$ and $\Phi' = g_{\mathscr{C}}(X'^n_a; s^n)$ and the random vector $X'^n_a$ is conditionally independent of $X_a^n$, i.e.,

$$\begin{aligned} & p_{X'^n_a, X_a^n|X_e^n,S^n}(x'^n_a, x_a^n|x_e^n, s^n) \\ & \quad = p_{X_a^n|X_e^n,S^n}(x'^n_a|x_e^n, s^n) p_{X_a^n|X_e^n,S^n}(x_a^n|x_e^n, s^n) \ . \end{aligned}$$

Step (b) follows from the convexity of $-\log(\cdot)$ and Jensen's inequality.

In Appendix D, we show that the expectation inside the logarithm in (58) is upper bounded as

$$\begin{aligned} & E_{\mathscr{C}}\left[\sum_{k,\phi} \tilde{V}_{s^n}(k, \phi|x_e^n) \Pr\Big((K', \Phi') = (k, \phi) | X_e^n = x_e^n, S^n = s^n, \mathscr{C}\Big)^{\alpha}\right] \\ & \qquad \leq \sum_{x_a^n} V_{s^n}(x_a^n|x_e^n)^{1+\alpha} + \frac{1}{|\mathcal{K}|^{\alpha} |\mathcal{M}|^{\alpha}} \ . \end{aligned} \tag{59}$$

We continue (58) by pulling out a term that depends only on $\log |\mathcal{K}||\mathcal{M}| = n(R_{\mathrm{M}} + R_{\mathrm{SK}})$ and taking the $\sum_{x_e^n} \tilde{Q}_{s^n}(x_e^n)$ into the logarithm through an application of Jensen's inequality.

$$\begin{aligned} & E_{\mathscr{C}}[H(K_a, \Phi | X_e^n, S^n = s^n, \mathscr{C})] \\ & \geq n(R_{\mathrm{M}} + R_{\mathrm{SK}}) \ldots \\ & - \frac{1}{\alpha} \log\left(1 + e^{n\alpha(R_{\mathrm{M}} + R_{\mathrm{SK}})} \sum_{x_e^n} \tilde{Q}_{s^n}(x_e^n) \sum_{x_a^n} V_{s^n}(x_a^n|x_e^n)^{1+\alpha}\right) \\ & \geq n(R_{\mathrm{M}} + R_{\mathrm{SK}}) - \frac{1}{\alpha} \exp\left(-n\left[F_0(\alpha, P_{s^n}) - \alpha(R_{\mathrm{M}} + R_{\mathrm{SK}})\right]\right) \end{aligned} \tag{60}$$

where for the second inequality we apply the relation $\log(1 + x) \leq x$ and let

$$F_0(\alpha, P_{s^n}) = -\frac{1}{n} \log\left(\sum_{x_e^n} \tilde{Q}_{s^n}(x_e^n) \sum_{x_a^n} V_{s^n}(x_a^n|x_e^n)^{1+\alpha}\right).$$

Using the memoryless property of the source, we further simplify $F_0(\alpha, P_{s^n})$

$$\begin{aligned} & F_0(\alpha, P_{s^n}) \\ & = -\frac{1}{n} \log\left(\prod_{i=1}^{n} \sum_{x_e} \tilde{Q}_{s_i}(x_e) \sum_{x_a} V_{s_i}(x_a|x_e)^{1+\alpha}\right) \\ & = -\frac{1}{n} \log\left(\prod_{s \in \mathcal{S}} \left[\sum_{x_e} \tilde{Q}_{s_i}(x_e) \sum_{x_a} V_{s_i}(x_a|x_e)^{1+\alpha}\right]^{np_S(s)}\right) \\ & = \sum_{s \in \mathcal{S}} P_{s^n}(s) \tilde{F}_0(\alpha, s) \end{aligned} \tag{61}$$

where

$$\tilde{F}_0(\alpha, s) = -\log\left(\sum_{x_e} \tilde{Q}_s(x_e) \sum_{x_a} V_s(x_a|x_e)^{1+\alpha}\right).$$

We complete the secrecy exponent proof by combining (55), (60), (61) and letting $P_{s^n} \to p_S$ as $n \to \infty$.

*3) Existence of a Deterministic Code for Both Exponential Reliability and Secrecy:* We have shown that there is a good code that achieves exponentially decay in error probability and that there is a good code that achieves exponentially decay in information leakage by characterizing their ensemble performances. To complete the proof, the last thing we need to show is the existence of a single code achieving both (reliability and secrecy) conditions.

To this end, note that $\Pr(\mathcal{E}_k) = E_{\mathscr{C}}[\Pr(\mathcal{E}_k|\mathscr{C})] = \sum_{\mathcal{C}} p_{\mathscr{C}}(\mathcal{C}) \Pr(\mathcal{E}_k|\mathcal{C})$. Let $\mathcal{A}_1$ denote the event that $\Pr(\mathcal{E}_k|\mathcal{C}) \geq 3\Pr(\mathcal{E}_k)$. By Markov's inequality, $\Pr(\mathcal{A}_1) \leq \frac{1}{3}$. Similarly, the key leakage can be expressed as $I(K_a; X_e^n, \Phi|S^n = s^n) = E_{\mathscr{C}}[I(K_a; X_e^n, \Phi|S^n = s^n, \mathscr{C})]$. Let $\mathcal{A}_2$ be the event that for some code $\mathcal{C}$ in the ensemble $I(K_a; X_e^n, \Phi|S^n = s^n, \mathcal{C}) \geq 3E_{\mathscr{C}}[I(K_a; X_e^n, \Phi|S^n = s^n, \mathscr{C})]$, so we have $\Pr(\mathcal{A}_2) \leq \frac{1}{3}$. From the union bound,

$$\begin{aligned}\Pr(\mathcal{A}_1^c \cap \mathcal{A}_2^c) &= 1 - \Pr(\mathcal{A}_1 \cup \mathcal{A}_2) \\ &\geq 1 - \Pr(\mathcal{A}_1) - \Pr(\mathcal{A}_2) \geq \frac{1}{3} \ ,\end{aligned}$$

so $\mathcal{A}_1^c \cap \mathcal{A}_2^c \neq \emptyset$. We complete the proof.

## D. Proof of Theorem 3

The properties of $\tilde{E}_0(\rho, s)$ are developed in [24, Theorem 2]. Here we show the properties of $\tilde{F}_0(\alpha, s)$.

It is easy to check $\tilde{F}_0(0, s) = 0$. Taking the derivative

$$\frac{\partial \tilde{F}_0(\alpha, s)}{\partial \alpha} = \frac{-\sum_{x_e} \tilde{Q}_s(x_e) \sum_{x_a} V_s(x_a|x_e)^{1+\alpha} \log V_s(x_a|x_e)}{\sum_{x_e} \tilde{Q}_s(x_e) \sum_{x_a} V_s(x_a|x_e)^{1+\alpha}}$$

we see $\frac{\partial \tilde{F}_0(\alpha,s)}{\partial \alpha} \geq 0$ for $\alpha \geq 0$. Thus, $\tilde{F}_0(\alpha, s)$ is an increasing non-negative function for $\alpha \geq 0$. Also, $\frac{\partial \tilde{F}_0(\alpha,s)}{\partial \alpha}$ equals $H(X_a|X_e, S = s)$ when $\alpha = 0$. To see $\tilde{F}_0(\alpha, s)$ is a concave function for $\alpha \geq 0$, let $\alpha_3 = \lambda\alpha_1 + (1-\lambda)\alpha_2$ where $0 < \lambda < 1$.

$$\begin{aligned}&\sum_{x_e} \tilde{Q}_s(x_e) \sum_{x_a} V_s(x_a|x_e)^{1+\alpha_3} \\ &= \sum_{x_e} \tilde{Q}_s(x_e) \left( \sum_{x_a} V_s(x_a|x_e)^{\lambda(1+\alpha_1)} V_s(x_a|x_e)^{(1-\lambda)(1+\alpha_2)} \right) \\ &\overset{\text{(i)}}{\leq} \sum_{x_e} \tilde{Q}_s(x_e) \left( \sum_{x_a} V_s(x_a|x_e)^{1+\alpha_1} \right)^{\lambda} \cdots \\ &\qquad \times \left( \sum_{x_a} V_s(x_a|x_e)^{1+\alpha_2} \right)^{1-\lambda} \\ &\overset{\text{(ii)}}{\leq} \left( \sum_{x_e} \tilde{Q}_s(x_e) \sum_{x_a} V_s(x_a|x_e)^{1+\alpha_1} \right)^{\lambda} \cdots \\ &\qquad \times \left( \sum_{x_e} \tilde{Q}_s(x_e) \sum_{x_a} V_s(x_a|x_e)^{1+\alpha_2} \right)^{1-\lambda} ,\end{aligned}$$

where (i) follows by Hölder's inequality:

$$\sum_i a_i b_i \leq \Big( \sum_i a_i^{1/\gamma} \Big)^{\gamma} \Big( \sum_i b_i^{1/(1-\gamma)} \Big)^{1-\gamma},$$

for $0 \leq \gamma \leq 1$ where we use $\gamma = \lambda$, and (ii) follows by a variant of Hölder's inequality [30, Problem 4.15]:

$$\sum_i p_i a_i b_i \leq \Big( \sum_i p_i a_i^{1/\gamma} \Big)^{\gamma} \Big( \sum_i p_i b_i^{1/(1-\gamma)} \Big)^{1-\gamma},$$

where $0 \leq \gamma \leq 1$, $\sum_i p_i = 1$, and again we select $\gamma = \lambda$. Taking $-\log(\cdot)$ on both sides, we get

$$\tilde{F}_0(\alpha_3, s) \geq \lambda \tilde{F}_0(\alpha_1, s) + (1-\lambda)\tilde{F}_0(\alpha_2, s) \ .$$

## E. Proof of Theorem 4

In the equal-SNR case ($\gamma_a = \gamma_b = \gamma$), the secret key rate achieved using an on-off excitation with parameter $\lambda$ is $\lambda I_{\mathrm{K}}(\gamma/\lambda)$ where $I_{\mathrm{K}}(x) = \log(1 + x^2/(1+2x))$, cf. (29)–(30). In (42) we use $\lambda_{\mathrm{c}}$ to denote the optimal on-off excitation parameter. Let $\bar{I}_{\mathrm{K}}(\gamma) = \lambda_{\mathrm{c}} I_{\mathrm{K}}(\gamma/\lambda_{\mathrm{c}})$. We note that $\bar{I}_{\mathrm{K}}(\gamma)$ is increasing and concave in $\gamma$. The latter follows from the observation made in Section IV-B that $I_{\mathrm{K}}(\gamma)$ is concave at high-SNR and due to the time-sharing. We now show that $\bar{I}_{\mathrm{K}}(\gamma)$ is the secret key capacity of this system.

We start by bounding the secret key capacity from above by Theorem 1. In the following $\mathcal{P}_S = \{p_S : E[|S|^2] \leq \mathsf{P}\}$ is the allowable set of excitation signals and $\gamma = \mathsf{P}/\sigma^2$ is the maximum achievable system SNR.

$$\begin{aligned}C_{\mathrm{K}} &\leq \max_{p_S \in \mathcal{P}_S} I(X_a; X_b|S) = \max_{p_S \in \mathcal{P}_S} \sum_{s \in \mathcal{S}} I(X_a; X_b|S = s) p_S(s) \\ &= \max_{p_S \in \mathcal{P}_S} \sum_{s \in \mathcal{S}} \left[ I_{\mathrm{K}}\left(\frac{|s|^2}{\sigma^2}\right) \right] p_S(s) \\ &\leq \max_{p_S \in \mathcal{P}_S} \sum_{s \in \mathcal{S}} \left[ \max_{\lambda, 0 \leq \lambda \leq 1} \lambda I_{\mathrm{K}}\left(\frac{|s|^2}{\sigma^2 \lambda}\right) \right] p_S(s) \\ &= \max_{p_S \in \mathcal{P}_S} \sum_{s \in \mathcal{S}} \left[ \bar{I}_{\mathrm{K}}\left(\frac{|s|^2}{\sigma^2}\right) \right] p_S(s) \\ &\leq \max_{p_S \in \mathcal{P}_S} \bar{I}_{\mathrm{K}}\left( \frac{\sum_{s \in \mathcal{S}} |s|^2 p_S(s)}{\sigma^2} \right) \\ &= \max_{p_S \in \mathcal{P}_S} \bar{I}_{\mathrm{K}}(\gamma_{p_S}) \leq \bar{I}_{\mathrm{K}}(\gamma).\end{aligned}$$

where the last inequality follows from the concavity of $\bar{I}_{\mathrm{K}}(\cdot)$ and $\gamma_{p_S}$ is the SNR achieved by distribution $p_S$.

## F. Proof of Theorem 6

To characterize the achievable error exponents of EDMSs with continuous random variables, in Appendix A we study the exponents achieved in a discretized (quantized) system and let the quantization get increasingly fine. In the limit the finite sum over the probability mass function (PMF) in (18) is replaced by an integral across a probability density function (PDF), and entropy in (27) is replaced by differential entropy.

The resulting form of the exponent is

$$E_{\mathrm{R}}(R_{\mathrm{SK}}) = \max_{p_S \in \mathcal{P}_S} \max_{0 \le \rho \le 1} \rho[h(X_a|S) - R_{\mathrm{SK}}] - E_0(\rho, p_S) \tag{62}$$

$$E_0(\rho, p_S) = \int_s p_S(s) \left[ \log \int_{x_b} Q_s(x_b) \ldots \left( \int_{x_a} W_s(x_a|x_b)^{\frac{1}{1+\rho}} dx_a \right)^{1+\rho} dx_b \right] ds \tag{63}$$

For the rest of this section we consider the special case of Theorem 6, namely when the input excitation signal is a constant signal, i.e., $p_S(s) = \delta(s - s_0)$ with $\frac{|s_0|^2}{\sigma^2} = \gamma$. First note that since $s = s_0$ with probability one,

$$E_0(\rho, p_S) = \log \int_{x_b} Q_{s_0}(x_b) \left( \int_{x_a} W_{s_0}(x_a|x_b)^{\frac{1}{1+\rho}} dx_a \right)^{1+\rho} dx_b \tag{64}$$

Next, due to the use of a constant excitation signal, $X_a$ and $X_b$ are jointly Gaussian complex random variables. Their distribution is

$$\mathcal{CN}\left( \mathbf{0}, \begin{bmatrix} \sigma_a^2 & \eta\sigma_a\sigma_b \\ \eta^*\sigma_a\sigma_b & \sigma_b^2 \end{bmatrix} \right)$$

where $\eta$ is the correlation coefficient. In the case of the signal model considered in Section II-C, cf. (8a)-(8b), one can verify the relations

$$\sigma_a^2 = \sigma_b^2 = (1+\gamma)\sigma^2, \qquad \eta = \frac{\gamma}{1+\gamma}.$$

The conditional PDF $W_{s_0}(x_a|x_b)$ is also complex Gaussian with distribution $\mathcal{CN}\left(m_{a|b}, \sigma_{a|b}^2\right)$ where

$$m_{a|b} = \frac{\eta\sigma_a}{\sigma_b} y = \frac{\gamma}{1+\gamma} y \ ,$$

$$\sigma_{a|b}^2 = (1-\eta^2)\sigma_a^2 = \frac{\sigma_a^2}{1+\gamma_{\mathrm{eq}}} \ ,$$

and $\gamma_{\mathrm{eq}} = \gamma^2/(1+2\gamma)$.

With these distributions, after some calculation it can show the inner integral of (64) is $(\pi\sigma_{a|b}^2)^{\frac{\rho}{1+\rho}}(1+\rho)$ and (64) simplifies to

$$E_0(\rho, p_S) = \rho \log(\pi\sigma_{a|b}^2) + (1+\rho)\log(1+\rho).$$

To find $E_{\mathrm{R}}(R_{\mathrm{SK}}, \gamma)$, we take a partial derivative with respect to $\rho$ of $\rho[h(X_a|S = s_0) - R_{\mathrm{SK}}] - E_0(\rho, p_S)$. Setting the result equal to zero yields

$$\begin{aligned} R_{\mathrm{SK}} &= h(X_a|S = s_0) - \frac{\partial E_0(\rho, p_S)}{\partial \rho} \\ &= \log(\pi e \sigma_a^2) - \log(\pi e \sigma_{a|b}^2(1+\rho)) \\ &= \log\left( \frac{1+\gamma_{\mathrm{eq}}}{1+\rho} \right) \end{aligned}$$

where $h(X_a|S = s_0) = \log(\pi e \sigma_a^2)$ since the variables are jointly complex Gaussians. Using the relation $I_{\mathrm{K}}(\gamma) = \log(1+\gamma_{\mathrm{eq}})$ we can solve for $\rho$ as

$$\rho = \exp[I_{\mathrm{K}}(\gamma) - R_{\mathrm{SK}}] - 1 \ . \tag{65}$$

Since $0 \le \rho \le 1$, the $\rho$ specified in (65) optimizes (62) only when $I_{\mathrm{K}}(\gamma) - \log 2 \le R_{\mathrm{SK}} \le I_{\mathrm{K}}(\gamma)$. We conclude that, when $\gamma_{\mathrm{eq}} \ge 1$, the error exponent expressed in terms of $\gamma$ falls into three regions. First, if $R_{\mathrm{SK}} \ge I_{\mathrm{K}}(\gamma)$ then

$$E_{\mathrm{R}}(R_{\mathrm{SK}}, \gamma) = 0.$$

Next, if $I_{\mathrm{c}}(\gamma) \le R_{\mathrm{SK}} < I_{\mathrm{K}}(\gamma)$, where $I_{\mathrm{c}}(\gamma) = \log\left(\frac{1+\gamma_{\mathrm{eq}}}{2}\right)$, then

$$E_{\mathrm{R}}(R_{\mathrm{SK}}, \gamma) = \rho[I_{\mathrm{K}}(\gamma) - R_{\mathrm{SK}} + 1] - (1+\rho)\log(1+\rho).$$

And, finally, if $0 \le R_{\mathrm{SK}} < I_{\mathrm{c}}(\gamma)$

$$E_{\mathrm{R}}(R_{\mathrm{SK}}, \gamma) = I_{\mathrm{K}}(\gamma) - R_{\mathrm{SK}} + 1 - 2\log 2.$$

When $0 \le \gamma_{\mathrm{eq}} < 1$, only the first two regions exist since $I_{\mathrm{c}}(\gamma) < 0$.

## VI. Conclusions

We investigate secret key generation from EDMS in this paper. We characterize secret key capacity and show the strong achievability of secrecy and reliability. This achievable region shows the tradeoff between secrecy and reliability. The results are applied to a case in which the EDMS is the output of Rayleigh fading channels when the participants transmit excitation signals to excite the channel randomness. We show that an on-off signal can achieve secret key capacity for all SNRs. The capacity-achieving on-off signal has a vanishing duty cycle when SNR approaches zero. An on-off excitation signal also achieves a higher error exponent than a uniform excitation signal. All these improvements due to choosing an input excitation distribution have a great impact on energy consumption in the low-SNR regime.

## Appendix A
## Reliability exponent of continuous random variables

To deal with continuous random variables, such as jointly Gaussian variables, we quantize the variables, apply (27), (17) – (18), and characterize the limit of increasingly fine quantization. Let $X_a^{\triangle}, X_b^{\triangle}$ be the quantized versions of continuous random variables $X_a$ and $X_b$ with uniform step size $\triangle$. Let $Q_s(x_a, x_b)$ denote the joint PDF indexed by $S = s$, i.e., $Q_s(x_a, x_b) = p_{X_a X_b|S}(x_a, x_b|s)$. Similarly, let $Q_s(x_b)$ and $W_s(x_a|x_b)$ denote its marginal and conditional distributions, respectively. Finally, let $Q_s^{\triangle}(i,j)$, $Q_s^{\triangle}(j)$ and $W_s^{\triangle}(i|j)$ be the corresponding quantized PMFs. By the mean value theorem,

$$\begin{aligned} Q_s^{\triangle}(j) W_s^{\triangle}(i|j) &= Q_s^{\triangle}(i,j) \\ &= \int_{x_b = j\triangle}^{(j+1)\triangle} \int_{x_a = i\triangle}^{(i+1)\triangle} Q_s(x_a, x_b) dx_a dx_b \\ &= Q_s(x_{a,i}, x_{b,j})\triangle^2 = Q_s(x_{b,j}) W_s(x_{a,i}|x_{b,j})\triangle^2 \ , \end{aligned}$$

for some $x_{a,i} \in (i\triangle, (i+1)\triangle)$ and $x_{b,j} \in (j\triangle, (j+1)\triangle)$. Applying (18) to the quantized variables gives

$$\begin{aligned}
&\tilde{E}_0^{\triangle}(\rho, s) \\
&= \log \sum_j \left( \sum_i \left[ Q_s(x_{b,j}) W_s(x_{a,i}|x_{b,j}) \triangle^{1+\rho} \triangle^{1-\rho} \right]^{\frac{1}{1+\rho}} \right)^{1+\rho} \\
&= \log \left( \triangle^{-\rho} \sum_j \triangle Q_s(x_{b,j}) \left( \sum_i \triangle \left[ W_s(x_{a,i}|x_{b,j}) \right]^{\frac{1}{1+\rho}} \right)^{1+\rho} \right) \\
&= -\rho \log \triangle + \ \ldots \\
&\qquad \log \left( \sum_j \triangle Q_s(x_{b,j}) \left( \sum_i \triangle W_s(x_{a,i}|x_{b,j})^{\frac{1}{1+\rho}} \right)^{1+\rho} \right) .
\end{aligned}$$

Substituting this into (27) for a fixed $s$ gives

$$\begin{aligned}
&\max_{0\le\rho\le 1} \rho \left( H(X_a^{\triangle}|S=s) - R_{\mathrm{SK}} \right) - \tilde{E}_0^{\triangle}(\rho, s) \\
&= \max_{0\le\rho\le 1} \rho \left[ H(X_a^{\triangle}|S=s) + \log\triangle - R_{\mathrm{SK}} \right] \\
&\quad - \log \left( \sum_j \triangle Q_s(x_{b,j}) \left( \sum_i \triangle W_s(x_{a,i}|x_{b,j})^{\frac{1}{1+\rho}} \right)^{1+\rho} \right) .
\end{aligned}$$

In the limit of increasingly fine quantization, i.e., as $\triangle$ approaches 0, $H(X_a^{\triangle}|S=s) + \log\triangle$ approaches $h(X_a|S=s)$, and the summation inside the logarithm approaches to an integral form. Thus, we can design a system whose reliability function can be made arbitrarily close to

$$E_{\mathrm{R}}(R_{\mathrm{SK}}) = \max_{p_S \in \mathcal{P}_S} \max_{0\le\rho\le 1} \rho[h(X_a|S) - R_{\mathrm{SK}}] - E_0(\rho, p_S) \ ,$$

where $E_0(\rho, p_S)$ is of the form (63).

## Appendix B
## Proof of Equation (47)

Consider the equivalent Gaussian channel model presented in (37)–(39). We can write the equation for the real part as

$$Y = X + W \ ,$$

where $X = \mathrm{Re}(\beta X_a)$, $Y = \mathrm{Re}(X_b)$ and $W = \mathrm{Re}(Z)$. We have

$$\begin{aligned}
X &\sim \mathcal{N}\left(0, \sigma_1^2\right), \quad \sigma_1^2 = \frac{1}{2} \frac{\mathsf{P}^2}{\mathsf{P} + \sigma_a^2} \\
W &\sim \mathcal{N}\left(0, \sigma_0^2\right), \quad \sigma_0^2 = \frac{1}{2} \left[ \left(\mathsf{P} + \sigma_b^2\right) - \left( \frac{\mathsf{P}^2}{\mathsf{P} + \sigma_a^2} \right) \right] .
\end{aligned}$$

Let $\mathcal{H}_0 = \{x : x \ge 0\}$ and $\mathcal{H}_1 = \{x : x < 0\}$. Then, the transition probability can be calculated to be

$$\begin{aligned}
\theta &= \Pr(Y \in \mathcal{H}_1 | X \in \mathcal{H}_0) = \frac{\Pr(Y \in \mathcal{H}_1, X \in \mathcal{H}_0)}{\Pr(X \in \mathcal{H}_0)} \\
&= 2 \int_{x\in\mathcal{H}_0} \Pr(Y \in \mathcal{H}_1 | X = x) p_X(x) dx \\
&= 2 \int_{x>0} \Pr(X + W < 0 | X = x) p_X(x) dx \\
&= 2 \int_{x>0} \frac{1}{2} \left[ 1 - \mathrm{erf}\left( \frac{x}{\sqrt{2\sigma_0^2}} \right) \right] p_X(x) dx \\
&= \int_{x>0} \left[ 1 - \mathrm{erf}\left( \frac{x}{\sqrt{2\sigma_0^2}} \right) \right] \frac{1}{\sqrt{2\pi\sigma_1^2}} \exp\left( \frac{-x^2}{2\sigma_1^2} \right) dx \\
&= \frac{1}{2} - \frac{1}{\sqrt{2\pi\sigma_1^2}} \int_{x>0} \mathrm{erf}\left( \frac{x}{\sqrt{2\sigma_0^2}} \right) \exp\left( \frac{-x^2}{2\sigma_1^2} \right) dx \\
&= \frac{1}{2} - \frac{1}{\pi} \arctan\left( \frac{\sigma_1}{\sigma_0} \right) .
\end{aligned}$$

The last equality follows from the formula

$$\int_0^\infty \mathrm{erf}(ax) e^{-b^2x^2} dx = \frac{1}{\sqrt{\pi} b} \arctan\left( \frac{a}{b} \right), \ \text{for } b \ne 0.$$

Finally, since $\frac{\sigma_1}{\sigma_0} = \sqrt{\gamma_{\mathrm{eq}}}$,

$$\theta = \frac{1}{2} - \frac{1}{\pi} \arctan\left( \sqrt{\gamma_{\mathrm{eq}}} \right) .$$

## Appendix C
## Error Exponent of DSBS

For a discrete memoryless source $(X, Y)$ with distribution $Q(y)W(x|y)$, in [24, Theorem 4] Gallager shows that the error exponent can be expressed as a generalized entropy function

$$\begin{aligned}
E_{\mathrm{R}} &= H(X_\rho Y_\rho \| XY) \\
&\triangleq \sum_{x,y} Q_\rho(y) W_\rho(x|y) \log \frac{Q_\rho(y) W_\rho(x|y)}{Q(y) W(x|y)}, \qquad (66)
\end{aligned}$$

where $0 \le \rho \le 1$ and $Q_\rho(y)$ and $W_\rho(x|y)$ are tilted distributions defined as

$$\begin{aligned}
Q_\rho(y) &= \frac{Q(y) \left( \sum_x W(x|y)^{1/(1+\rho)} \right)^{(1+\rho)}}{\sum_y Q(y) \left( \sum_x W(x|y)^{1/(1+\rho)} \right)^{(1+\rho)}} \ , \\
W_\rho(x|y) &= \frac{W(x|y)^{1/(1+\rho)}}{\sum_x W(x|y)^{1/(1+\rho)}} \ .
\end{aligned}$$

The message rate is parameterized by $\rho$ and has the form

$$R_{\mathrm{M}} = H(X_\rho | Y_\rho) \triangleq \sum_y -Q_\rho(y) W_\rho(x|y) \log W_\rho(x|y).$$

For a binary sources with $Q(y) = (\frac{1}{2}, \frac{1}{2})$ and symmetric transition probability $W(x|y) = \theta$ if $x \ne y$ and $W(x|y) = 1 - \theta$ if $x = y$. $Q_\rho(y)$ is uniformly $(\frac{1}{2}, \frac{1}{2})$ distributed and

$$W_\rho(x|y) = \frac{\theta^{1/(1+\rho)}}{\theta^{1/(1+\rho)} + (1-\theta)^{1/(1+\rho)}} \triangleq \tau \ \text{ if } x \ne y.$$

If $0 \leq \rho \leq 1$ then $\theta \leq \tau \leq \frac{\sqrt{\theta}}{\sqrt{\theta}+\sqrt{1-\theta}}$ and $R_{\mathrm{M}}$ is in the range

$$H_B(\theta) \leq R_{\mathrm{M}} \leq H_B\left(\frac{\sqrt{\theta}}{\sqrt{\theta}+\sqrt{1-\theta}}\right).$$

The corresponding error exponent (66) can be simplified to

$$\begin{aligned} E_{\mathrm{R}}(R_{\mathrm{M}}) &= H(X_\rho Y_\rho \| XY) \\ &= \tau \log \frac{\tau}{\theta} + (1-\tau)\log\frac{1-\tau}{1-\theta} \\ &= T_\theta(\tau) - H_B(\tau)\ , \end{aligned}$$

where $H_B(\cdot)$ is the binary entropy function and $T_\theta(\tau) \triangleq -\tau\log(\theta) - (1-\tau)\log(1-\theta)$. When $R_{\mathrm{M}} < H_B(\theta)$ one finds that $E_{\mathrm{R}}(R_{\mathrm{M}}) = 0$ (by choosing $\rho = 0$). When $R_{\mathrm{M}} > H_B\left(\frac{\sqrt{\theta}}{\sqrt{\theta}+\sqrt{1-\theta}}\right)$, $E_{\mathrm{R}}(R_{\mathrm{M}})$ is maximized by $\rho = 1$. The resulting error exponent is

$$E_{\mathrm{R}}(R_{\mathrm{M}}) = R_{\mathrm{M}} - 2\log(\sqrt{\theta}+\sqrt{1-\theta}).$$

## Appendix D
## Proof of Equation (59)

We adopt the notation in (56). For fixed $k, \phi$ and $\mathcal{C}$, the probability of a collision event is

$$\begin{aligned} &\Pr\Big((K', \Phi') = (k, \phi) | X_e^n = x_e^n, S^n = s^n, \mathscr{C} = \mathcal{C}\Big) \\ &= \sum_{k', \phi'} \tilde{V}_{s^n, \mathcal{C}}(k', \phi' | x_e^n) \mathbf{1}[k = k', \phi = \phi'] \end{aligned}$$

The derivation is shown in next page. The step (67) is our first application of Jensen's Inequality to the term in brackets $(\cdot)^\alpha$ since the sum over $k, \phi$ is a sum over the probability mass function $\mathbf{1}[k, \phi | x_a^n, s^n, \mathcal{C}]$ (cf. (57) for the definition of this indicator function). We recall that $x_a^n$, $s^n$, and $\mathcal{C}$ are all fixed for this inner sum, the last being fixed by the outer expectation over $\mathscr{C}$. In (68) we split the sum over $x_a'^n$ into two terms and distribute the sums over $k'$ and $\phi'$. We then apply the "sifting" property of the indicator function that selects out a single term from the sum. We next apply the inequality $(x+y)^\alpha \leq x^\alpha + y^\alpha$, for $0 \leq \alpha \leq 1$ to get (69). In (70) we note that the first term is not a function of $\mathcal{C}$. We next note that the function $f(x) = x^\alpha$ is concave if $0 \leq \alpha \leq 1$. We use this to move both the sum over $x_a^n$ and the expectation over codebooks inside the function, a step justified by Jensen's Inequality. In (71) we apply the uniformly random design of the binning functions. Since $x_a^n \neq x_a'^n$ *for every* term in the sum, each of the indicator functions equals the (fixed) pair $(k, \phi)$ with equal probability and independently. Thus, the probability that *both* equal $(k, \phi)$ is the square (by the independence) of the reciprocal of the number of possibilities (by the uniformity). Noting that $V_{s^n}(x_a^n | x_e^n) V_{s^n}(x_a'^n | x_e^n)$ is a well defined (conditional) probability mass function and that we are missing one term in the double sum, we get the (actually strict) upper bounding by unity in (72).

## Acknowledgement

The authors would like to thank Vincent Y. F. Tan for the discussions that helped to improve the content of the work.

## References

[1] T. Chou, A. Sayeed, and S. Draper, "Minimum energy per bit for secret key acquisition over multipath wireless channels," in *Workshop on Inform. Theory Applications*, (San Diego, CA), Jan. 2009.

[2] T. Chou, A. Sayeed, and S. Draper, "Minimum energy per bit for secret key acquisition over multipath wireless channels," in *Information Theory, 2009. ISIT 2009. IEEE International Symposium on*, (Seoul, Korea), June 2009.

[3] C. E. Shannon, "Communication theory of secrecy systems," *Bell Systems Tech. Journal*, vol. 28, pp. 656–715, Oct. 1949.

[4] A. Wyner, "The wire-tap channel," *The Bell Systems Technical Journal*, vol. 54, pp. 1355 – 1387, Oct. 1975.

[5] I. Csiszár and J. Körner, "Broadcast channels with confidential messages," *Information Theory, IEEE Transactions on*, vol. 24, pp. 339–348, May 1978.

[6] U. M. Maurer, "Secret key agreement by public discussion from common information," *Information Theory, IEEE Transactions on*, vol. 39, pp. 733–742, May 1993.

[7] R. Ahlswede and I. Csiszár, "Common randomness in information theory and cryptography Part I: Secret sharing," *Information Theory, IEEE Transactions on*, vol. 39, pp. 1121–1132, July 1993.

[8] D. Slepian and J. Wolf, "Noiseless coding of correlated information sources," *Information Theory, IEEE Transactions on*, vol. 19, pp. 471–480, July 1973.

[9] D. Tse and P. Viswanath, *Fundamentals of Wireless Communication*. Cambridge University Press, 2005.

[10] C. A. Balanis, *Antenna Theory: Analysis and Design*. New York: Wiley, 2nd ed., 1997.

[11] S. Verdu, "On channel capacity per unit cost," *Information Theory, IEEE Transactions on*, vol. 36, pp. 1019–1030, Sep 1990.

[12] C. Ye, A. Reznik, and Y. Shah, "Extracting secrecy from jointly gaussian random variables," in *Information Theory, 2006 IEEE International Symposium on*, pp. 2593–2597, July 2006.

[13] S. Nitinawarat, "Secret key generation for correlated gaussian sources," in *Proceedings of the 45th annual Allerton Conference*, 2007.

[14] A. A. Hassan, W. E. Stark, J. E. Hersheyc, and S. Chennakeshu, "Cryptographic key agreement for mobile radio," *Digital Signal Processing*, vol. 6, pp. 207–212, Oct. 1996.

[15] A. Sayeed and A. Perrig, "Secure wireless communications: Secret keys through multipath," in *ICASSP 2008. IEEE Int. Conf. on*, (Las Vegas, NV), pp. 3013–3016, 2008.

[16] R. Wilson, D. Tse, and R. A. Scholtz, "Channel identification: Secret sharing using reciprocity in ultrawideband channels," *Information Forensics and Security, IEEE Transactions on*, vol. 2, pp. 364–375, Sep. 2007.

[17] C. Ye, A. Reznik, G. Sternberg, and Y. Shah, "On the secrecy capabilities of itu channels," in *Vehicular Technology Conference, 2007. VTC-2007 Fall. 2007 IEEE 66th*, pp. 2030–2034, Sep. 2007.

[18] C. Ye, S. Mathur, A. Reznik, Y. Shah, W. Trappe, and N. Mandayam, "Information-theoretically secret key generation for fading wireless channels," *Information Forensics and Security, IEEE Transactions on*, vol. 5, pp. 240 –254, June 2010.

[19] J. Wallace, C. Chen, and M. Jensen, "Key generation exploiting mimo channel evolution: Algorithms and theoretical limits," in *Antennas and Propagation, 2009. EuCAP 2009. 3rd European Conference on*, pp. 1499–1503, Mar. 2009.

[20] N. Patwari, J. Croft, S. Jana, and S. K. Kasera, "High-rate uncorrelated bit extraction for shared secret key generation from channel measurements," *Mobile Computing, IEEE Trans. on*, vol. 9, pp. 17–30, Jan. 2010.

[21] M. Bloch, J. Barros, M. Rodrigues, and S. McLaughlin, "Wireless information-theoretic security," *Information Theory, IEEE Transactions on*, vol. 54, pp. 2515–2534, June 2008.

[22] A. Khisti, S. Diggavi, and G. Wornell, "Secret-key generation with correlated sources and noisy channels," in *Information Theory, 2008. ISIT 2008. IEEE International Symposium on*, pp. 1005–1009, July 2008.

[23] U. Maurer and S. Wolf, "Information-theoretic key agreement: from weak to strong secrecy for free," in *EUROCRYPT'00: Proc. of the 19th Int. Conf. on Theory and application of cryptographic techniques*, pp. 351–368, Springer-Verlag, 2000.

[24] R. G. Gallager, "Source coding with side information and universal coding," *M.I.T. LIDS-P-937*, Sep. 1979.

[25] C. Bennett, G. Brassard, C. Crepeau, and U. Maurer, "Generalized privacy amplification," *Information Theory, IEEE Transactions on*, vol. 41, pp. 1915 –1923, Nov. 1995.

[26] M. Hayashi, "Exponential decreasing rate of leaked information in

$$
\begin{aligned}
& E_{\mathscr{C}}\left\{\sum_{k,\phi}\tilde{V}_{s^n}(k,\phi|x_e^n)\Pr\left((K',\Phi')=(k,\phi)|X_e^n=x_e^n,S^n=s^n,\mathscr{C}\right)^{\alpha}\right\} \\
&= E_{\mathscr{C}}\left\{\sum_{k,\phi}\left[\tilde{V}_{s^n,\mathscr{C}}(k,\phi|x_e^n)\left(\sum_{k',\phi'}\tilde{V}_{s^n,\mathscr{C}}(k',\phi'|x_e^n)\mathbf{1}[k=k',\phi=\phi']\right)^{\alpha}\right]\right\} \\
&= E_{\mathscr{C}}\left\{\sum_{k,\phi}\left[\left(\sum_{x_a^n}V_{s^n}(x_a^n|x_e^n)\mathbf{1}[k,\phi|x_a^n,s^n,\mathscr{C}]\right)\left(\sum_{k',\phi'}\tilde{V}_{s^n,\mathscr{C}}(k',\phi'|x_e^n)\mathbf{1}[k=k',\phi=\phi']\right)^{\alpha}\right]\right\} \\
&= E_{\mathscr{C}}\left\{\sum_{x_a^n}V_{s^n}(x_a^n|x_e^n)\left[\sum_{k,\phi}\mathbf{1}[k,\phi|x_a^n,s^n,\mathscr{C}]\left(\sum_{k',\phi'}\tilde{V}_{s^n,\mathscr{C}}(k',\phi'|x_e^n)\mathbf{1}[k=k',\phi=\phi']\right)^{\alpha}\right]\right\} \\
&= E_{\mathscr{C}}\left\{\sum_{x_a^n}V_{s^n}(x_a^n|x_e^n)\left[\sum_{k,\phi}\mathbf{1}[k,\phi|x_a^n,s^n,\mathscr{C}]\left(\sum_{k',\phi'}\tilde{V}_{s^n,\mathscr{C}}(k',\phi'|x_e^n)\mathbf{1}[k=k',\phi=\phi']\right)\right]^{\alpha}\right\} && (67) \\
&= E_{\mathscr{C}}\left\{\sum_{x_a^n}V_{s^n}(x_a^n|x_e^n)\left[\sum_{k,\phi}\mathbf{1}[k,\phi|x_a^n,s^n,\mathscr{C}]\left(\sum_{k',\phi'}\left(\sum_{x_a'^n}V_{s^n}(x_a'^n|x_e^n)\mathbf{1}[k',\phi'|x_a'^n,s^n,\mathscr{C}]\right)\mathbf{1}[k=k',\phi=\phi']\right)\right]^{\alpha}\right\} \\
&= E_{\mathscr{C}}\left\{\sum_{x_a^n}V_{s^n}(x_a^n|x_e^n)\left[\sum_{x_a'^n}V_{s^n}(x_a'^n|x_e^n)\left(\sum_{k,\phi}\sum_{k',\phi'}\mathbf{1}[k,\phi|x_a^n,s^n,\mathscr{C}]\mathbf{1}[k',\phi'|x_a'^n,s^n,\mathscr{C}]\mathbf{1}[k=k',\phi=\phi']\right)\right]^{\alpha}\right\} \\
&= E_{\mathscr{C}}\left\{\sum_{x_a^n}V_{s^n}(x_a^n|x_e^n)\left[V_{s^n}(x_a^n|x_e^n)+\sum_{x_a'^n\neq x_a^n}V_{s^n}(x_a'^n|x_e^n)\left(\sum_{k,\phi}\mathbf{1}[k,\phi|x_a^n,s^n,\mathscr{C}]\mathbf{1}[k,\phi|x_a'^n,s^n,\mathscr{C}]\right)\right]^{\alpha}\right\} && (68) \\
&\le E_{\mathscr{C}}\left\{\sum_{x_a^n}V_{s^n}(x_a^n|x_e^n)\left\{[V_{s^n}(x_a^n|x_e^n)]^{\alpha}+\left[\sum_{x_a'^n\neq x_a^n}V_{s^n}(x_a'^n|x_e^n)\left(\sum_{k,\phi}\mathbf{1}[k,\phi|x_a^n,s^n,\mathscr{C}]\mathbf{1}[k,\phi|x_a'^n,s^n,\mathscr{C}]\right)\right]^{\alpha}\right\}\right\} && (69) \\
&\le \sum_{x_a^n}V_{s^n}^{1+\alpha}(x_a^n|x_e^n)+\left[E_{\mathscr{C}}\left\{\sum_{x_a^n}V_{s^n}(x_a^n|x_e^n)\sum_{x_a'^n\neq x_a^n}V_{s^n}(x_a'^n|x_e^n)\left(\sum_{k,\phi}\mathbf{1}[k,\phi|x_a^n,s^n,\mathscr{C}]\mathbf{1}[k,\phi|x_a'^n,s^n,\mathscr{C}]\right)\right\}\right]^{\alpha} && (70) \\
&= \sum_{x_a^n}V_{s^n}^{1+\alpha}(x_a^n|x_e^n)+\left[\sum_{x_a^n}V_{s^n}(x_a^n|x_e^n)\sum_{x_a'^n\neq x_a^n}V_{s^n}(x_a'^n|x_e^n)\left(\sum_{k,\phi}E_{\mathscr{C}}\left\{\mathbf{1}[k,\phi|x_a^n,s^n,\mathscr{C}]\mathbf{1}[k,\phi|x_a'^n,s^n,\mathscr{C}]\right\}\right)\right]^{\alpha} \\
&= \sum_{x_a^n}V_{s^n}^{1+\alpha}(x_a^n|x_e^n)+\left[\sum_{x_a^n}V_{s^n}(x_a^n|x_e^n)\sum_{x_a'^n\neq x_a^n}V_{s^n}(x_a'^n|x_e^n)\left(\sum_{k,\phi}\frac{1}{(|\mathcal{K}||\mathcal{M}|)^2}\right)\right]^{\alpha} && (71) \\
&= \sum_{x_a^n}V_{s^n}^{1+\alpha}(x_a^n|x_e^n)+\frac{1}{|\mathcal{K}|^{\alpha}|\mathcal{M}|^{\alpha}}\left[\sum_{x_a^n}\sum_{x_a'^n\neq x_a^n}V_{s^n}(x_a^n|x_e^n)V_{s^n}(x_a'^n|x_e^n)\right]^{\alpha} \\
&\le \sum_{x_a^n}V_{s^n}^{1+\alpha}(x_a^n|x_e^n)+\frac{1}{|\mathcal{K}|^{\alpha}|\mathcal{M}|^{\alpha}}. && (72)
\end{aligned}
$$

universal random privacy amplification," *Information Theory, IEEE Trans. on*, vol. 57, pp. 3989–4001, June 2011.
[27] C. H. Bennett, G. Brassard, and J.-M. Robert, "Privacy amplification by public discussion," *SIAM J. Comput.*, vol. 17, Apr. 1988.
[28] L. Zheng, D. N. C. Tse, and M. Medard, "Channel coherence in the low-SNR regime," *Information Theory, IEEE Transactions on*, vol. 53, pp. 976 –997, Mar. 2007.
[29] V. Raghavan, G. Hariharan, and A. Sayeed, "Capacity of sparse multipath channels in the ultra-wideband regime," *Selected Topics in Signal Processing, IEEE Journal of*, vol. 1, pp. 357 –371, Cct. 2007.
[30] R. G. Gallager, *Information theory and reliable communication*. New York: Wiley, 1968.

**Tzu-Han Chou** (S'06) was born in Taipei, Taiwan. He received the B.S. degree in electrical engineering and the M.S. degree in communication engineering, respectively, from National Taiwan University, and Ph.D degree in electrical and computer engineering at the University of Wisconsin, Madison.

He joined the Computer and Communications Research Labs, Industrial Technology Research Institute, Hsinchu, Taiwan and joined Sunplus Technology, Hsinchu, Taiwan, where he worked on GPRS/WCDMA baseband design. Currently, he is a senior system engineer with Qualcomm Inc, San Diego, CA. His research interests are in the area of wireless communication, information theory, and physical layer security.

**Stark C. Draper** (S'99-M'03) is Assistant Professor of Electrical and Computer Engineering at the University of Wisconsin, Madison. He received the M.S. and Ph.D. degrees in Electrical Engineering and Computer Science from the Massachusetts Institute of Technology (MIT), and the B.S. and B.A. degrees in Electrical Engineering and History, respectively, from Stanford University.

Before moving to Wisconsin, Dr. Draper worked at the Mitsubishi Electric Research Laboratories (MERL) in Cambridge, MA. He held postdoctoral positions in the Wireless Foundations, University of California, Berkeley, and in the Information Processing Laboratory, University of Toronto, Canada. He has worked at Arraycomm, San Jose, CA, the C. S. Draper Laboratory, Cambridge, MA, and Ktaadn, Newton, MA. His research interests include communication and information theory, error-correction coding, statistical signal processing and optimization, security, and application of these disciplines to computer architecture and semiconductor device design.

Dr. Draper has received an NSF CAREER Award, the UW ECE Gerald Holdridge Teaching Award, the MIT Carlton E. Tucker Teaching Award, an Intel Graduate Fellowship, Stanford's Frederick E. Terman Engineering Scholastic Award, and a U.S. State Department Fulbright Fellowship.

**Akbar M. Sayeed** (S'89-M'97-SM'02) is Professor of Electrical and Computer Engineering at the University of Wisconsin-Madison. He received the B.S. degree from the University of Wisconsin-Madison in 1991, and the M.S. and Ph.D. degrees from the University of Illinois at Urbana-Champaign in 1993 and 1996, all in Electrical and Computer Engineering.

He was a postdoctoral fellow at Rice University from 1996 to 1997. His research interests include wireless communications, statistical signal processing, multi-dimensional communication theory, time-frequency analysis, information theory, and applications in wireless communication and sensor networks.

Dr. Sayeed is a recipient of the Robert T. Chien Memorial Award (1996) for his doctoral work at Illinois, the NSF CAREER Award (1999), the ONR Young Investigator Award (2001), and the UW Grainger Junior Faculty Fellowship (2003). He is currently serving on the signal processing for communications and networking technical committee of the IEEE Signal Processing Society. Dr. Sayeed also served as an Associate Editor for the IEEE Signal Processing Letters from 1999-2002, and as the technical program co-chair for the 2007 IEEE Statistical Signal Processing Workshop and the 2008 IEEE Communication Theory Workshop.